\documentclass[journal]{IEEEtran}
\usepackage{cite}
\usepackage[dvips]{graphicx}
\usepackage[cmex10]{amsmath}
\usepackage{color}
\usepackage{bm}
\usepackage{theorem}
\usepackage{amscd}
\usepackage{amssymb}

{\theorembodyfont{\normalfont}
\theoremheaderfont{\normalfont\it}
\newtheorem{thm}{ Theorem}
\newtheorem{dfn}[thm]{ Definition}
\newtheorem{lmm}[thm]{ Lemma}

\newtheorem{crl}[thm]{ Corollary}

\newtheorem{prp}[thm]{ Proposition}
}

{\theorembodyfont{\normalfont}
\theoremheaderfont{\normalfont\it}
\newtheorem{prf}{ Proof:}}

{\theorembodyfont{\normalfont}
\theoremheaderfont{\normalfont\it}
}

{\theorembodyfont{\normalfont}
\theoremheaderfont{\normalfont\it}
}

{\theorembodyfont{\normalfont}
\theoremheaderfont{\normalfont\it}
}

{\theorembodyfont{\normalfont}
\theoremheaderfont{\normalfont\it}
}

{\theorembodyfont{\normalfont}
\theoremheaderfont{\normalfont\it}
\newtheorem{rmk}{ Remark.}}

\newcommand{\bra}[1]{\mbox{$\langle#1|$}}

\newcommand{\ket}[1]{\mbox{$|#1\rangle$}}

\newcommand{\outpro}[2]{\mbox{$\ket{#1}\!\bra{#2}$}}

\newcommand{\proj}[1]{\mbox{$\ket{#1}\!\bra{#1}$}}

\newcommand{\norm}[1]{\|#1\|}

\newcommand{\alg}[1]{\begin{align}#1\end{align}}
\newcommand{\rarrow}{\rightarrow}

\newcommand{\nn}{\nonumber}
\newcommand{\ca}[1]{{\mathcal #1}}

\newcommand{\bthm}[1]{\begin{thm}\label{thm:#1}}
\newcommand{\ethm}{\end{thm}}

\newcommand{\blmm}[1]{\begin{lmm}\label{lmm:#1}}
\newcommand{\elmm}{\end{lmm}}
\newcommand{\rLmm}[1]{Lemma \ref{lmm:#1}}

\newcommand{\bdfn}[1]{\begin{dfn}\label{dfn:#1}}
\newcommand{\edfn}{\end{dfn}}
\newcommand{\rdfn}[1]{Definition \ref{dfn:#1}}
\newcommand{\bprp}[1]{\begin{prp}\label{prp:#1}}
\newcommand{\eprp}{\end{prp}}

\newcommand{\rPrp}[1]{Proposition \ref{prp:#1}}
\newcommand{\bcrl}[1]{\begin{crl}\label{crl:#1}}
\newcommand{\ecrl}{\end{crl}}

\newcommand{\bprf}{\begin{prf}}
\newcommand{\eprf}{\end{prf}}
\newcommand{\brmk}{\begin{rmk}}
\newcommand{\ermk}{\end{rmk}}
\newcommand{\laeq}[1]{\label{eq:#1}}
\newcommand{\req}[1]{(\ref{eq:#1})}

\newcommand{\QED}{\hfill$\blacksquare$}
\newcommand{\lsec}[1]{\label{sec:#1}}
\newcommand{\rsec}[1]{\ref{sec:#1}}
\newcommand{\rSec}[1]{Section \ref{sec:#1}}

\newcommand{\rApp}[1]{Appendix \ref{app:#1}}

\newcommand{\bitem}{\begin{itemize}}
\newcommand{\entem}{\end{itemize}}
\newcommand{\benum}{\begin{enumerate}}
\newcommand{\ennum}{\end{enumerate}}
\newcommand{\bb}{\mathbb}
\newcommand{\otm}{\otimes}
\newcommand{\opl}{\oplus}
\newcommand{\vro}{\varrho}
\newcommand{\rFig}[1]{Figure \ref{fig:#1}}

\begin{document} 

\title{Operational Resource Theory of Non-Markovianity}

\author{Eyuri Wakakuwa
\thanks{E. Wakakuwa is with the Department of Complexity Science and Engineering, Graduate School of Frontier Sciences, The University of Tokyo, Japan (email: e.wakakuwa@edu.k.u-tokyo.ac.jp).} 
}

\maketitle

\begin{abstract}

In order to analyze non-Markovianity of tripartite quantum states from a resource theoretical viewpoint, 
we introduce a class of quantum operations performed by three distant parties, and investigate an operational resource theory (ORT) induced it. 
A tripartite state is a free state if and only if it is a quantum Markov chain. 
We prove monotonicity of functions such as the conditional quantum mutual information, intrinsic information, squashed entanglement, a generalization of entanglement of purification, and the relative entropy of recovery. 
The ORT has five bound sets, each of which corresponds to one of the monotone functions.
We introduce a task of ``non-Markovianity dilution'', and prove that the optimal rate for the task, namely the ``non-Markovianity cost'', is bounded from above by entanglement of purification in the case of pure states. 
We also propose a classical resource theory of non-Markovianity.

\end{abstract}


\hfill

\section{Introduction}

Markovianity of a physical system composed of three subsystems is a property that the first and the second subsystems are statistically independent when conditioned by the third one. In classical information theory, the system is modeled by random variables $X$, $Y$ and $Z$, which take values in finite sets $\ca X$, $\ca Y$ and $\ca Z$, respectively, according to a joint probability distribution $\{p(x,y,z)\}_{(x,y,z)\in{{\ca X}\times{\ca Y}\times{\ca Z}}}$. The three systems are referred to as being a Markov chain in the order of $X\rightarrow Z\rightarrow Y$ if the probability is decomposed as $p(x,y,z)=p(x,z)p(y|z)=p(x|z)p(y|z)p(z)$, or equivalently, if the conditional mutual information $I(X:Y|Z)$ is equal to zero. In analogy, a {\it quantum Markov chain} is defined as a tripartite quantum states for which the conditional quantum mutual information is zero \cite{hayden04}. Classical probability distributions and quantum states for which the conditional mutual information is {\it not} equal to zero have the property of {\it non-Markovianity}.

Non-Markov classical and quantum states can be used as a resource for several information processing tasks that are impossible without using such states: Ref.~\cite{gisin2000linking,maurer1999unconditionally,renner2003new} proved that classical random variables that obey non-Markov distribution may be used in secret key agreement protocols, and Ref.~\cite{sharma2017conditional} proved that non-Markov quantum states are used as a resource for a task called the {\it conditional quantum one-time pad}. This situation resembles the one that entangled quantum states shared between distant parties can be used as a resource for tasks that cannot be accomplished without using entanglement (see e.g. \cite{plenio07,horodecki2009quantum}). Hence, it would be natural to apply concepts and tools, developed in entanglement theory\cite{plenio07,horodecki2009quantum}, to an analysis of non-Markovianity of classical and quantum states.

In this paper, we apply the concept of {\it operational resource theory} (ORT)\cite{horodecki2013quantumness,brandao2015reversible,gour2017quantum,anshu2017quantifying}, which originates in entanglement theory, for analyzing non-Markovianity of tripartite quantum states from an operational point of view. We consider a scenario in which three distant parties Alice, Bob and Eve perform operations on their systems by communicating classical messages, quantum messages and applying local operations.  We restrict the class $\Omega$ of free operations to be one consisting of (i) local operations by Alice and Bob, (ii) local reversible operations by Eve, (iii) broadcasting of classical messages by Alice and Bob, (iv) quantum communication from Alice and Bob to Eve, and their compositions. We analyze an ORT induced by $\Omega$. It turns out that a tripartite quantum state is a free state if and only if it is a quantum Markov chain. Thus the obtained ORT is regarded as an {\it operational resource theory of non-Markovianity}. 

One of the principal goals of an ORT in general is to identify 1) necessary and sufficient condition for one state to be convertible to another by free operations ({\it single-shot convertibility}), and 2) the optimal ratio of number of copies at which one state is asymptotically convertible to another by free operations ({\it asymptotic convertibility}). These are, however, often a highly complex problem as in the case for an ORT of multipartite entanglement (see e.g. \cite{plenio07,guhne2009entanglement,amico2008entanglement,horodecki2009quantum,walter2016multi}). In such cases, a key milestone would be to identify 3) subsets of states that are closed under free operations and tensor product ({\it bound sets}), and 4) real-valued functions of states that are monotonically nonincreasing under free operations ({\it monotones}).

 In this paper, we mainly address 3) and 4) above for an ORT induced by $\Omega$. We also address 1) by considering particular examples of qubit systems, and 2) by introducing and analyzing a task of {\it non-Markovianity dilution}, in which copies of a unit state is transformed to copies of another state by a free operation. We note that our approach is different from that of \cite{rivas2014quantum}, which addresses quantification of non-Markovianity of {\it processes}.

This paper is organized as follows. \rSec{settings} provides the setting of the problem. 
In \rSec{freestate}, we prove that the conditional quantum mutual information is monotonically nonincreasing under free operations, and that a state is a free state if and only if it is a quantum Markov chain. 
In \rSec{NMmonot}, we identify functions that are monotonically non-increasing under free operations, including a quantum analog of intrinsic information\cite{gisin2000linking,maurer1999unconditionally,renner2003new}, a generalization of entanglement of purification\cite{terhal2002entanglement}, the squashed entanglement\cite{christandl2002quantum} and the relative entropy of recovery\cite{sesh14}.  In \rSec{boundset}, we find five bound sets, each of which have a clear correspondence to the monotone functions mentioned above, and analyze inclusion relations among them. In \rSec{duality}, we prove that the monotone functions and the bound sets are connected with each other by a type of ``duality'', which is analogous to the duality of CQMI for four-partite pure states\cite{berta2015renyi}. Examples of single-shot convertibility are provided in Section \ref{sec:unitres}. In \rSec{disdil}, we introduce a task of {\it non-Markovianity dilution}, and prove that the {\it non-Markovianity cost}, namely the optimal achievable rate in non-Markovianity dilution, is bounded from above by the entanglement of purification for the case of pure states. In \rSec{classical}, we consider a {\it classical} resource theory of non-Markovianity. Conclusions are given in \rSec{discussion}. Some of the proofs of the main results are provided in appendices.\\

{\it Notations.}  
A system composed of two subsystems $A$ and $B$ is denoted by $AB$. 
We abbreviate $|\psi\rangle^A\otimes|\phi\rangle^B$ as $|\psi\rangle^A|\phi\rangle^B$. 
We denote $(M^A\otimes I^B)\ket{\psi}^{AB}$ as $M\ket{\psi}^{AB}$.
When ${\mathcal E}$ is a quantum operation on $A$, we denote $({\mathcal E}\otimes{\rm id}^B)(\rho^{AB})$ as $({\mathcal E}^A\otimes{\rm id}^B)(\rho^{AB})$ or ${\mathcal E}(\rho^{AB})$.  
For $\rho^{AB}$, $\rho^{A}$ represents ${\rm Tr}_B[\rho^{AB}]$. We denote $|\psi\rangle\!\langle\psi|$ simply as $\psi$. 
The von Neumann entropy of a state $\rho^A$ is interchangeably denoted by $S(\rho^A)$ and $S(A)_\rho$.  $\log{x}$ represents the base $2$ logarithm of $x$.

\hfill

\section{Settings}\lsec{settings}

\begin{figure}[t]
\begin{center}
\includegraphics[bb={0 0 564 370}, scale=0.4]{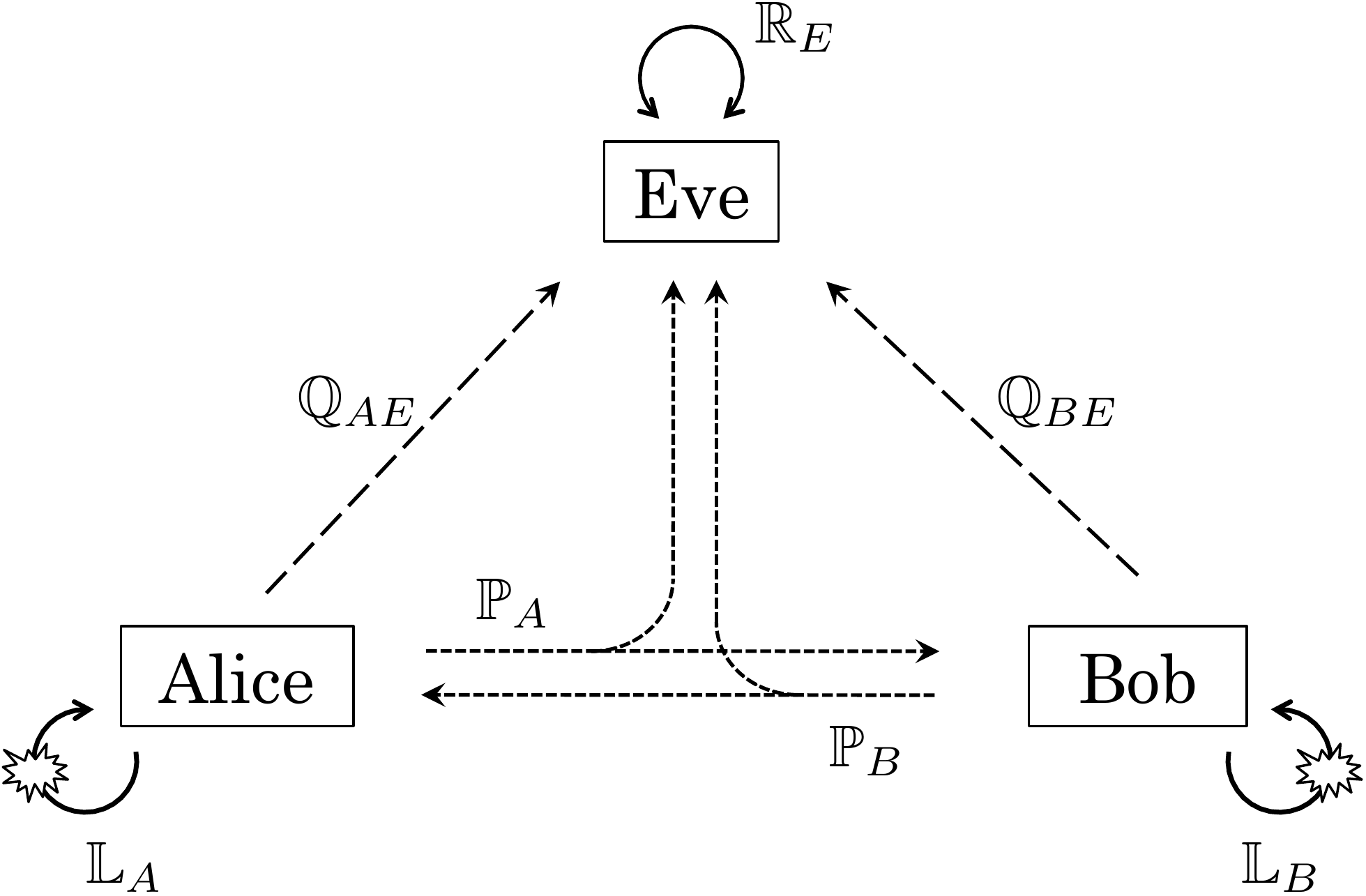}
\end{center}
\caption{The classes of operations that comprises free operations are depicted. Any operation in $\Omega$ is represented as a composition of operations in the classes depicted in this figure.}
\label{fig:freeoperations}
\end{figure}

Suppose three distant parties Alice, Bob and Eve have quantum systems $A$, $B$ and $E$, respectively. A quantum state on system $ABE$ is specified by finite-dimensional Hilbert spaces ${\ca H}^A$, ${\ca H}^B$, ${\ca H}^E$ and a density operator on ${\ca H}^A\otimes{\ca H}^B\otimes{\ca H}^E$. We denote the set of all quantum states on $ABE$ by ${\ca S}_{\rm all}$. An operation on system $S\in\{A,B,E\}$ is specified by a linear CPTP map ${\ca E}$ from ${\ca S}({\ca H}^S)$ to ${\ca S}({\ca H}'^S)$, where ${\ca H}^S$ and ${\ca H}'^S$ are Hilbert spaces corresponding to the input and the output of ${\ca E}$, respectively.

Consider the following classes of operations performed by Alice, Bob and Eve (see \rFig{freeoperations}):
\alg{
{\bb L}_{A}&\text{: local quantum operations by Alice}\nn\\
{\bb L}_{B}&\text{: local quantum operations by Bob}\nn\\
{\bb R}_E&\text{: local reversible quantum operation by Eve}\nn\\
{\bb P}_A&\text{: broadcasting of classical messages by Alice}\nn\\
{\bb P}_B&\text{: broadcasting of classical messages by Bob}\nn\\
{\bb Q}_{AE}&\text{: quantum communication from Alice to Eve}\nn\\
{\bb Q}_{BE}&\text{: quantum communication from Bob to Eve}\nn
}
By ``reversible'', we mean that for any operation ${\ca V}\in{\mathbb R}_E$, there exists an operation ${\ca V}^*$ by Eve such that ${\ca V}^*\circ{\ca V}$ is the identity operation on ${\ca S}({\ca H}_{{\rm In}({\ca V})})$, with ${\ca H}_{{\rm In}({\ca V})}$ denoting the Hilbert space corresponding to the input of $\ca V$. We require that Eve cannot refuse to receive anything that is sent to her in the above operations, i.e., quantum communication from Alice in ${\mathbb Q}_{AE}$, one from Bob in ${\mathbb Q}_{BE}$, classical messages from Alice in ${\mathbb P}_A$ and one from Bob in ${\mathbb P}_B$. This condition, together with the reversibility of Eve's operation, imposes a strong restriction on what the three parties can perform by the above classes of operations. We denote by $\Omega$ the set of operations that can be represented as a composition of operations belonging to the above classes.

In this paper, we consider a scenario in which Alice, Bob and Eve are only allowed to perform operations in $\Omega$. We analyze conditions under which a state on $ABE$ is convertible to another by an operation in $\Omega$. That is, we consider an operational resource theory induced by $\Omega$.
Due to the condition of reversibility of Eve's operations, it is too restrictive to define convertibility of a state $\rho_1$ to $\rho_2$ by the existence of an operation ${\ca E}\in\Omega$ such that ${\ca E}(\rho_1)=\rho_2$. Thus we relax the definition of state convertibility as follows.

\bdfn{zconvertibility}
{\it A state $\rho_1$ is convertible to $\rho_2$ under $\Omega$} if there exists an operation ${\ca E}\in\Omega$ and a reversible operation $\ca V$ on $E$ such that ${\ca E}(\rho_1^{ABE})={\ca V}(\rho_2^{ABE})$.
\edfn

It should be noted that, if we ignore Eve's system, ${\bb Q}_{AE}$ and ${\bb Q}_{BE}$ are regarded as subclasses of ${\bb L}_{A}$ and ${\bb L}_{B}$, respectively, and ${\bb P}_A$ and ${\bb P}_A$ as equivalent to classical communication between Alice and Bob. Hence any operation in $\Omega$ can be identified with local operations and classical communication (LOCC) between Alice and Bob, if we ignore Eve.

\hfill

\section{Basic Properties}\lsec{freestate}

In this section, we analyze basic properties of an operational resource theory (ORT) induced by $\Omega$. We prove that the degree of non-Markovianity of quantum states, measured by the conditional quantum mutual information, is monotonically nonincreasing under $\Omega$. We also prove that a state is a free state if and only if it is a quantum Markov chain. Thus an ORT proposed in this paper is regarded as that of non-Markovianity. The ``maximally non-Markovian state'', from which any other state of a given dimension is generated by an operation in $\Omega$, is identified.

\subsection{Monotonicity of Conditional Quantum Mutual Information}

The {\it conditional quantum mutual information (CQMI)} $I(A:B|E)_\rho$ of a tripartite quantum state $\rho$ on system $ABE$ is defined by
\alg{
&I(A:B|E)_\rho\nn\\
&\quad\quad:=S(AE)_\rho\!+\!S(BE)_\rho\!-\!S(ABE)_\rho\!-\!S(E)_\rho,\nn
}
where $S$ is the von Neumann entropy. The CQMI is nonnegative due to the strong subadditivity of the von Neumann entropy\cite{lieb1973proof}, and has operational meanings in terms e.g. of quantum state redistribution\cite{devetak2008exact,yard2009optimal,ye2008quantum}, deconstruction protocols\cite{berta2016deconstruction} and the conditional quantum one-time pad\cite{sharma2017conditional}. For simplicity, we denote $I(A:B|E)_\rho$ by $I_M(\rho)$. The following lemma states that $I_M$ is monotonically nonincreasing under $\Omega$.

\blmm{monotiM}
For any $\rho\in{\ca S}_{\rm all}$ and ${\ca E}\in\Omega$, we have $I_M(\rho)\geq I_M({\ca E}(\rho))$.
\elmm

\bprf
It suffices to prove that $I_M$ is monotonically nonincreasing under any class of operations that comprises $\Omega$.

\benum
\item {\it Monotonicity under ${\mathbb L}_A$ and ${\mathbb L}_B$}:
For any $\rho\in{\ca S}_{\rm all}$, ${\ca E}\in{\mathbb L}_A$ and ${\ca E}'\in{\mathbb L}_A$, we have
\alg{
I(A:B|E)_{\rho}\geq I(A:B|E)_{({\ca E}\otm{\ca E}')(\rho)},\nn
}
due to the data processing inequality for the conditional quantum mutual information. 

\item {\it Monotonicity under ${\mathbb R}_E$}:
For any ${\ca V}\in{\mathbb R}_E$, we have 
\alg{
I(A:BE)_\rho&\geq I(A:BE)_{{\ca V}(\rho)}\nn\\
&\geq I(A:BE)_{({\ca V}^*\circ{\ca V})(\rho)}\nn\\
&=I(A:BE)_\rho,\nn
}
due to the data processing inequality. Hence we obtain 
\alg{
I(A:BE)_\rho =I(A:BE)_{{\ca V}(\rho)}.\nn
}
In the same way, we also have
\alg{
I(A:E)_\rho =I(A:E)_{{\ca V}(\rho)}.\nn
}
Thus we have
\alg{
I(A:B|E)_\rho&=I(A:BE)_\rho-I(A:E)_\rho\nn\\
&=I(A:BE)_{{\ca V}(\rho)}-I(A:E)_{{\ca V}(\rho)}\nn\\
&=I(A:B|E)_{{\ca V}(\rho)},\nn
}
which implies the monotonicity (actually the invariance) of $I_M$ under ${\mathbb R}_E$.

\item{\it Monotonicity under ${\mathbb P}_A$ and ${\mathbb P}_B$}:
The state before broadcasting of classical message by Alice is represented by a density operator
\alg{
\rho_i=\sum_mp_m\proj{m}^{C_{\!A}}\otimes\rho_m^{ABE},\label{eq:rhohati}
} 
where $C_A$ is a register possessed by Alice, $\{p_m\}_m$ is a probability distribution, $\{|m\rangle\}_m$ is a set of orthonormal pure states, and $\rho_m$ is a density operator for each $m$. In the same way, the state after broadcasting is represented by
\alg{
\rho_f=\sum_mp_m\proj{m}^{C_{\!B}}\otimes\proj{m}^{C_{\!E}}\otimes\rho_m^{ABE},\label{eq:rhohatf}
} 
where $C_B$ and $C_E$ are registers possessed by Bob and Eve, respectively. We have
\alg{
I(C_AA:B|E)_{\rho_i}&=I(C_A:B|E)_{\rho_i}+I(A:B|EC_A)_{\rho_i}\nn\\
&\geq I(A:B|EC_A)_{\rho_i}\nn\\
&=\sum_mp_mI(A:B|E)_{\rho_m}\nn\\
&=I(A:C_BB|C_EE)_{\rho_f},\nn
}
which implies monotonicity under ${\mathbb P}_A$. The monotonicity under ${\mathbb P}_B$ follows along the same line.

\item{\it Monotonicity under ${\mathbb Q}_{AE}$ and ${\mathbb Q}_{BE}$}: 
Let $Q$ be a quantum system that is transmitted from Alice to Eve. For any quantum state $\rho$ on $ABEQ$, we have
\alg{
I(QA:B|E)_{\rho}
&=I(Q:B|E)_{\rho}+I(A:B|EQ)_{\rho}\nn\\
&\geq I(A:B|EQ)_{\rho},\label{eq:onakaitaii}
}
which implies monotonicity under ${\mathbb Q}_{AE}$. The monotonicity under ${\mathbb Q}_{BE}$ follows along the same line.

\ennum
\hfill$\blacksquare$
\eprf

\subsection{Quantum Markov Chains are Free States}\lsec{tashikani}

A quantum Markov chain is defined as a tripartite quantum state for which the conditional quantum mutual information (CQMI) is zero\cite{hayden04}. For a tripartite system composed of systems $A$, $B$ and $E$, the condition is represented as
\begin{eqnarray}
I(A:B|E)_\rho=0.\laeq{markovianity}
\end{eqnarray}
We denote the set of quantum Markov chains satisfying \req{markovianity} by ${\ca S}_{\rm Markov}$. Ref.~\cite{hayden04} proved that Equality (\ref{eq:markovianity}) is equivalent to the condition that there exists a quantum operation ${\mathcal R}:E\rightarrow BE$ satisfying 
\begin{eqnarray}
\rho^{ABE}={\mathcal R}(\rho^{AE}),\nn
\end{eqnarray}
as well as to the condition that there exists a linear isometry $\Gamma$ from $E$ to $E_0E_LE_R$ such that
\begin{eqnarray}
\Gamma\rho^{ABE}\Gamma^\dagger=\sum_{j\in{\ca J}}p_j\proj{j}^{E_0}\otm\sigma_j^{AE_L}\otimes\tau_j^{BE_R}.\laeq{decomposability}
\end{eqnarray} 
Here, $\{p_j\}_{j\in{\ca J}}$ is a probability distribution, $\{|j\rangle\}_{j\in{\ca J}}$ is an orthonormal basis of $E_0$, and $\sigma_j$ and $\tau_j$ are quantum states on composite systems $AE_L$ and $BE_R$, respectively, for each $j$. 

In an ORT, a state is called a {\it free state} if it can be generated from scratch by a free operation. That is, a state $\sigma\in{\ca S}_{\rm all}$ is called {\it a free state under $\Omega$} if, for any $\rho\in{\ca S}_{\rm all}$, there exists ${\ca E}\in\Omega$ such that ${\ca E}(\rho)=\sigma$. Due to the following proposition, an ORT induced by $\Omega$ is regarded as that of of non-Markovianity.

\bprp{freestate}
A state $\sigma^{ABE}$ is a free state under $\Omega$ if and only if $\sigma\in{\ca S}_{\rm Markov}$.
\eprp

\bprf
To prove the ``if'' part, consider the following procedure: 1. Alice generates a random variable $J$ which takes values in $\ca J$ according to a probability distribution $p_j$, 2. Alice broadcasts $J$ to Bob and Eve, 3. Eve records $J$ on her register, 3. Alice locally prepares a state $\sigma_j^{AE_L}$ and sends $E_L$ to Eve, 4. Bob locally prepares a state $\sigma_j^{BE_R}$ and sends $E_R$ to Eve, and 5. Alice and Bob discards $J$. It is straightforward to verify that any state in the form of \req{decomposability} can be generated by this protocol. 

To prove the ``only if'' part, suppose that $\sigma'\in{\ca S}_{\rm all}$ is a free state. By definition, for any $\sigma\in{\ca S}_{\rm Markov}$ there exists an operation ${\ca E}\in\Omega$ such that ${\ca E}(\sigma)=\sigma'$. From \rLmm{monotiM}, it follows that $0=I_M(\sigma)\geq I_M(\sigma')$, which yields $I_M(\sigma')=0$ and thus $\sigma'\in{\ca S}_{\rm Markov}$.\QED
\eprf

In an ORT, it would be natural to require that the set of free states is closed under tensor product \cite{horodecki2013quantumness,brandao2015reversible,gour2017quantum}. The following lemma states that this condition is satisfied in an ORT induced by $\Omega$.

\blmm{tensorMarkov}
For any $\sigma_1,\sigma_2\in{\ca S}_{\rm Markov}$, we have $\sigma_1\otimes\sigma_2\in{\ca S}_{\rm Markov}$.
\elmm

\bprf
It is straightforward to verify that for any $\sigma_1\in{\ca S}_{\rm Markov}$ on $A_1B_1E_1$ and $\sigma_2\in{\ca S}_{\rm Markov}$ on $A_2B_2E_2$ we have
\alg{
&I(A_1A_2:B_1B_2|E_1E_2)_{\sigma_1\otm\sigma_2}\nn\\
&\quad=I(A_1:B_1|E_1)_{\sigma_1}+I(A_2:B_2|E_2)_{\sigma_2}=0,\nn
}
which implies $\sigma_1\otimes\sigma_2\in{\ca S}_{\rm Markov}$.\QED
\eprf

\subsection{Maximally non-Markovian state}

Let $|\Phi_d\rangle$ be a $d$-dimensional maximally entangled state defined by 
\alg{
|\Phi_d\rangle:=\frac{1}{\sqrt{d}}\sum_{k=1}^d\ket{k}\ket{k}.\nn
}
and consider a state $\Phi_{{\rm I},d}\in{\ca S}_{\rm all}$ defined by
\alg{
\ket{\Phi_{{\rm I},d}}^{ABE}:=|\Phi_d\rangle^{AB}\ket{0}^E.\nn
}
Suppose that $A$ and $B$ are $d$-dimensional quantum systems. It is straightforward to verify that any state $\rho\in{\ca S}_{\rm all}$ is created from $\Phi_{{\rm I},d}$ by the following operation, which is an element of $\Omega$: 1. Alice locally prepares a state $\rho^{ABE}$, 2. Alice sends system $E$ to Eve, and 3. Alice teleports system $B$ to Bob by using $\Phi_{{\rm I},d}$ as a resource. Therefore, $\Phi_{{\rm I},d}$ is regarded as a $d$-dimensional ``maximally non-Markovian state'' on $ABE$. Note that the classical message transmitted by Alice in Step 3 is decoupled from the state obtained at the end of the protocol.

\subsection{Nonconvexity}

An ORT is said to be {\it convex} if the set of free states is convex, or equivalently, if any probabilistic mixture of free states is also a free state. Most of the ORTs proposed so far satisfies convexity (see \cite{brandao2015reversible,gour2017quantum,anshu2017quantifying} and the references therein), except that of non-Gaussianity\cite{braunstein2005quantum}. We show that an ORT induced by $\Omega$ is another example of nonconvex ORTs.

Consider states $\rho_1,\rho_2\in{\ca S}_{\rm all}$ defined by
\alg{
&\rho_0:=\proj{0}^A\otm\proj{0}^B\otm\proj{0}^E,\nn\\
&\rho_1:=\proj{1}^A\otm\proj{1}^B\otm\proj{0}^E,\nn
}
respectively. These states apparently satisfy $\rho_1,\rho_2\in{\ca S}_{\rm Markov}$. On the contrary, any probabilistic mixture of the two states, which takes the form of
\alg{
{\bar\rho}(\lambda):=\left(\lambda\proj{00}^{AB}+(1-\lambda)\proj{11}^{AB}\right)\otm\proj{0}^E\nn\\
(0<\lambda<1),\nn
}
does not belong to ${\ca S}_{\rm Markov}$ because 
\alg{
I_M({\bar\rho}(\lambda))=-\lambda\log{\lambda}-(1-\lambda)\log{(\lambda)}>0.\nn
}
Hence ${\ca S}_{\rm Markov}$ is not a convex set.


\hfill

\section{Monotones}\label{sec:NMmonot}

In this section, we introduce several functions of tripartite quantum states that are monotonically nonincreasing under $\Omega$. We will refer to these functions as {\it non-Markovianity monotones}, analogously to entanglement monotones for multipartite quantum states \cite{vidal2000entanglement,plenio07}. Proofs of the monotonicity are provided in Appendix \ref{app:monotII}.

\subsection{CQMI-based Monotones}\label{sec:CQMImonotone}

We introduce three non-Markovianity monotones based on the conditional quantum mutual information. The first one is a quantum mechanical generalization of {\it intrinsic information} \cite{gisin2000linking,maurer1999unconditionally,renner2003new,christandl2002quantum}, defined as
\alg{
I_{\downarrow}(\rho):=\inf_{{\ca T}}I(A:B|E)_{{\ca T}(\rho)}\nn
}
where infimum is taken over all operations ${\ca T}$ on $E$. The socond one is defined as
\alg{
I_{\downarrow}^*(\rho):=\inf_{\varrho}I(A:B|F)_{\varrho},\nn
}
where the infimum is taken over all extensions $\varrho$ of $\rho$ on $ABEF$, i.e., over all state $\varrho$ satisfying
\alg{
\varrho^{ABE}:={\rm Tr}_{F}[\varrho^{ABEF}]=\rho^{ABE}.\nn
}

Entanglement of bipartite reduced state $\rho^{AB}={\rm Tr}_E[\rho^{ABE}]$ is quantified by an entanglement measure $E^{A:B}$ for bipartite quantum states. The monotonicity of $E^{A:B}$ under $\Omega$ simply follows from the fact that any operation in $\Omega$ can be identified with LOCC between Alice and Bob, if we ignore Eve. We may choose for $E^{A:B}$ an arbitrary entanglement monotone \cite{vidal2000entanglement,plenio07} for bipartite quantum states. In this paper, we adopt the {\it squashed entanglement} \cite{christandl2002quantum,christandl04} defined by 
\alg{
E_{sq}(\tau):=\frac{1}{2}\inf_{\omega}I(A:B|R)_\omega\nn
}
for a bipartite state $\tau$ on $AB$, where the infimum is taken over all extensions $\omega^{ABR}$ of $\tau^{AB}$. Note that $E_{sq}(\tau)=0$ if and only if $\tau$ is a separable state. In the following, we will use the notation
\alg{
I_{sq}(\rho):=2E_{sq}({\rm Tr}_E[\rho^{ABE}])\nn
}
for $\rho\in{\ca S}_{\rm all}$.

\subsection{Generalized Entanglement of Purification}\label{sec:gentpure}

Consider a state $\rho\in{\ca S}_{\rm all}$ and suppose we split system $E$ to a composite system $E_AE_B$ by applying a linear isometry $\ca W$ from $E$ to $E_AE_B$. The entanglement of state ${\ca W}(\varrho)$ between $AE_A$ and $BE_B$ is quantified by the squashed entanglement. Let $F$ be an arbitrary quantum system, and $\varrho$ be a quantum state on $ABEF$ satisfying ${\rm Tr}_{F}[\varrho^{ABEF}]=\rho^{ABE}$. Noting that ${\rm Tr}_{F}[{\ca W}(\varrho)]={\ca W}(\rho)$, the squashed entanglement of ${\ca W}(\varrho)$ between $AE_A$ and $BE_B$ is given by
\alg{
2E_{sq}({\ca W}(\varrho))=\inf_{\varrho}I(AE_A:BE_B|F)_{{\ca W}(\varrho)},\nn
}
where the infimum is taken over all extensions $\varrho^{ABEF}$ of $\rho^{ABE}$. By taking the infimum over all splitting of $E$, we define
\alg{
J_\downarrow^*(\rho):=\inf_{\ca W}\inf_{\varrho}I(AE_A:BE_B|F)_{{\ca W}(\varrho)}.\nn
}
This quantity could be regarded as a generalization of the {\it entanglement of purification} introduced in \cite{terhal2002entanglement}. 

Let $F_A$ and $F_B$ be arbitrary quantum systems, and let $|\phi_\rho\rangle$ be a state on $ABEF_AF_B$ that is a purification of $\rho^{ABE}$, i.e., ${\rm Tr}_{F_AF_B}[|\phi_\rho\rangle\!\langle\phi_\rho|]=\rho^{ABE}$. The squashed entanglement of state $\phi_\rho$ between $AF_A$ and $BF_B$ is given by
\alg{
2E_{sq}(\phi_\rho)=\inf_{{\ca T}}I(AF_A:BF_B|E)_{{\ca T}(\phi_\rho)},\nn
}
where the infimum is taken over all operations $\ca T$ on $E$. Taking the infimum over all purifications of $\rho^{ABE}$, we define
\alg{
J_\downarrow(\rho):=\inf_{\phi_\rho}\inf_{{\ca T}}I(AF_A:BF_B|E)_{{\ca T}(\phi_\rho)}.\nn
}

Due to the data processing inequality for the CQMI, we have
\alg{
J_\downarrow(\rho)&=\inf_{\ca W}\inf_{\varrho}I(AE_A:BE_B|F)_{{\ca W}(\varrho)}\nn\\
&\geq \inf_{\phi_\rho}\inf_{{\ca T}}I(A:B|E)_{{\ca T}(\phi_\rho)}\nn\\
&=\inf_{{\ca T}}I(A:B|E)_{{\ca T}(\rho)}\nn\\
&=I_\downarrow(\rho)\laeq{hoshino1}
}
as well as
\alg{
J_\downarrow^*(\rho)&=\inf_{\ca W}\inf_{\varrho}I(AE_A:BE_B|F)_{{\ca W}(\varrho)}\nn\\
&\geq \inf_{\ca W}\inf_{\varrho}I(A:B|F)_{{\ca W}(\varrho)}\nn\\
&=\inf_{\varrho}I(AE_A:BE_B|F)_{{\ca W}(\rho)}\nn\\
&=I_\downarrow^*(\rho).\laeq{hoshino2}
}

\subsection{Relative Entropy of Recovery}\label{sec:owmonot}

Let $\Omega^\rightarrow$ be the set of operations that can be represented as a composition of operations belonging to ${\mathbb L}_{A}$, ${\mathbb L}_{B}$, ${\mathbb R}_E$, ${\mathbb P}_A$, ${\mathbb Q}_A$ and ${\mathbb Q}_B$ (but not to ${\mathbb P}_B$). Instead of $\Omega$, we may consider an ORT induced by $\Omega^\rightarrow$. It follows from the proof of \rPrp{freestate} (see \rSec{tashikani}) that a state $\sigma^{ABE}$ is a free state under $\Omega^\rightarrow$ if and only if $\sigma\in{\ca S}_{\rm Markov}$. 

The {\it relative entropy of recovery} \cite{sesh14} (see Remark 6 therein) is defined as 
\alg{
D_{\rm rec}^B(\rho):=\inf_{\ca R}D(\rho^{ABE}\|{\ca R}(\rho^{AE})),\nn
}
where the infimum is taken over all quantum operations $\ca R$ from $E$ to $BE$. The regularized version of the above function has an operational meaning in the context of quantum hypothesis testing\cite{cooney2016operational}. We prove in Appendix \ref{app:monotrer} that $D_{\rm rec}^B$ is monotonically nonincreasing under $\Omega^\rightarrow$. It is left open whether $D_{\rm rec}^B$ is a monotone under $\Omega$, or equivalently, whether it is monotonically nonincreasing under ${\mathbb P}_B$.

\hfill

\section{Bound sets}\label{sec:boundset}

In an operational resource theory (ORT), a set of states that is closed under free operations and tensor product is called a {\it bound set}. That is, a set ${\ca S}\in{\ca S}_{\rm all}$ is a {\it bound set under $\Omega$} if, for any $\rho,\sigma\in{\ca S}$ and ${\ca E}\in\Omega$, we have $\rho\otimes\sigma\in{\ca S}$ and ${\ca E}(\rho)\in{\ca S}$. In this section, we will show that an ORT induced by $\Omega$ has five bound sets in addition to ${\ca S}_{\rm Markov}$. Based on the equivalence between separability of bipartite states and Markovianity of its extension, we analyze inclusion relations among those bound sets.

\subsection{Definitions}\label{sec:defbound}

One of the five bound sets is the set of states that can be transformed to a quantum Markov chain by an (possibly irreversible) operation by Eve, i.e.,
\alg{
{\ca S}_I:=\{\rho\in{\ca S}_{\rm all}\:|\:\exists{\ca T}\in{\mathbb L}_E\text{ s.t. }{\ca T}(\rho)\in{\ca S}_{\rm Markov}\},\label{eq:midomi}
}
where we denote by ${\mathbb L}_E$ the sets of quantum operations on $E$. Let $F$ be an ancillary quantum system, and denote by ${\ca S}_{\rm Markov}^{A:B|F}$ the set of quantum Markov chains on $ABF$. Another bound set is the set of states on $ABE$ that has an extension on $ABEF$ such that the reduced state $ABF$ is a quantum Markov chain, namely,
\alg{
&{\ca S}_I^*:=\{\rho\in{\ca S}_{\rm all}\:|\:\exists\vro^{ABEF}\text{ s.t. }\nn\\
&\quad\quad\quad\quad\quad\quad\quad\varrho^{ABE}=\rho^{ABE},\;\varrho^{ABF}\in{\ca S}_{\rm Markov}^{A:B|F}\}.\nn
}
Since any operation in $\Omega$ is regarded as a LOCC between Alice and Bob by ignoring Eve, the set of states on $ABE$ that are separable between $A$ and $B$ when we trace out $E$ is also a bound set:
\alg{
{\ca S}_{\rm sep}:=\{\rho\in{\ca S}_{\rm all}\:|\:\rho^{AB}\text{ is separable}\}.\nn
}
By definition, it is straightforward to verify that the two bound sets defined above have a clear correspondence to the non-Markovianity monotones introduced in \rSec{NMmonot}. Namely, we have
\alg{
&I_\downarrow^*(\rho)=0\text{ for any }\rho\in{\ca S}_I^*\laeq{idasm},\\
&I_\downarrow(\rho)=0\text{ for any }\rho\in{\ca S}_I,\laeq{idasmi}\\
&I_{sq}(\rho)=0\text{ for any }\rho\in{\ca S}_{\rm sep}.\laeq{idasmii}
}

Suppose we split system $E$ to a composite system $E_AE_B$ by applying a linear isometry $\ca W$ from $E$ to $E_AE_B$. Depending on the state $\rho\in{\ca S}_{\rm all}$, we may choose $\ca W$ so that the state after splitting, namely ${\ca W}(\rho)$, is a separable state between $AE_A$ and $BE_B$. The set of such states is proved to be a bound set:
\alg{
&{\ca S}_J^*:=\{\rho\in{\ca S}_{\rm all}\:|\:\exists{\ca W}:E\rightarrow E_AE_B\text{ s.t. }\nn\\
&\quad\quad\quad\quad\quad\quad\quad\quad\quad\quad{\ca W}(\rho^{ABE})\in{\ca S}_{\rm sep}^{AE_A:BE_B}\}.\nn
}
Here, we denoted by ${\ca S}_{\rm sep}^{AE_A:BE_B}$ the set of separable states between $AE_A$ and $BE_B$. Let $F_A$ and $F_B$ be arbitrary quantum systems, and let $|\phi_\rho\rangle$ be a state on $ABEF_AF_B$ that is a purification of $\rho^{ABE}$, i.e., ${\rm Tr}_{F_AF_B}[|\phi_\rho\rangle\!\langle\phi_\rho|]=\rho^{ABE}$. By properly choosing $|\phi_\rho\rangle$, the reduced state on $ABF_AF_B$ may be a separable state between $AF_A$ and $BF_B$. We define
\alg{
&{\ca S}_J:=\{\rho\in{\ca S}_{\rm all}\:|\:\exists|\phi_\rho\rangle^{ABEF_AF_B}\text{ s.t. }\nn\\
&\quad\quad\quad\quad\quad\quad\phi_\rho^{ABE}=\rho^{ABE},\;\phi_\rho^{ABF_AF_B}\in{\ca S}_{\rm sep}^{AF_A:BF_B}\}.\nn
}
We will prove in the next subsection that similar relations as \req{idasm}-\req{idasmii} hold between $J_\downarrow^*$, $J_\downarrow$ and ${\ca S}_J^*$, ${\ca S}_J$, respectively (see \req{idasmiii} and \req{idasmiiii}).

\begin{figure}[t]
\begin{center}
\includegraphics[bb={0 0 674 410}, scale=0.37]{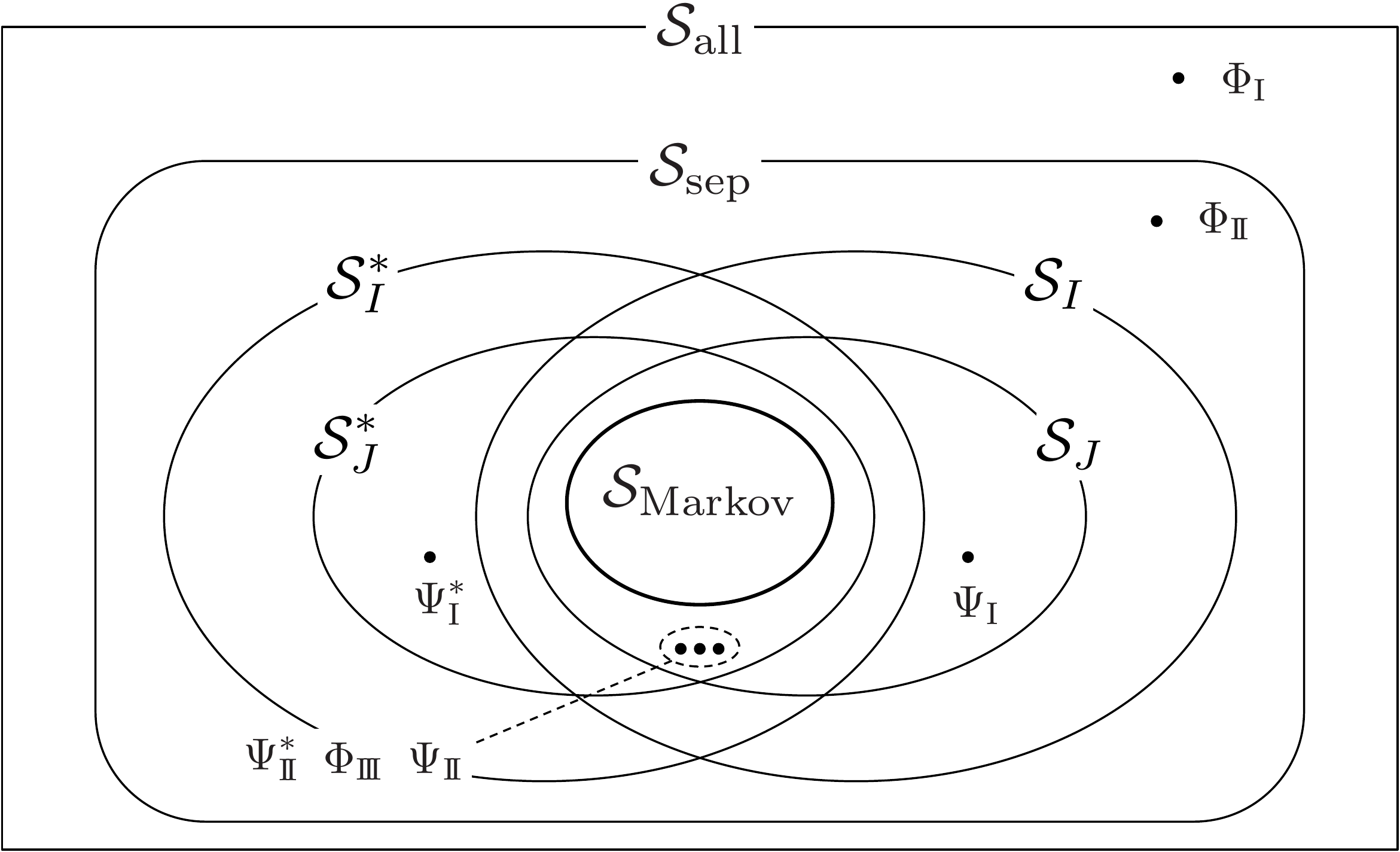}
\end{center}
\caption{Inclusion relations among the bound sets are depicted. The black dots in the subsets represent examples of states defined in \rSec{unitres}.}
\label{fig:inclusion}
\end{figure}

A proof that the sets ${\ca S}_I$, ${\ca S}_I^*$, ${\ca S}_J$, ${\ca S}_J^*$ and ${\ca S}_{\rm sep}$ are bound sets is separately provided in Appendix \ref{app:monotII}. We will prove in \rSec{inclusion} that the bound sets introduced above are connected with each other by the following inclusion relations (see \rFig{inclusion}):
\alg{
&{\ca S}_J\cap{\ca S}_J^*\supsetneq{\ca S}_{\rm Markov},\label{eq:kudakero}\\
&{\ca S}_I\cup{\ca S}_I^*\subsetneq{\ca S}_{\rm sep},\label{eq:moushikomi}\\
&{\ca S}_J\nsubseteq{\ca S}_I^*,\;{\ca S}_I\nsupseteq{\ca S}_J^*,\label{eq:atatte}\\
&{\ca S}_J\subseteq{\ca S}_I,\;{\ca S}_I^*\supseteq{\ca S}_J^*.\laeq{kukukku}
}
It is left open whether relations \req{kukukku} are strict.

\subsection{Equivalence of Separability and Markovianity}\lsec{eqsepmark}

The condition that a bipartite quantum state is separable is equivalent to the condition that there exists a quantum Markov chain that is an extension of the state, and to the condition that its purification is mapped to a quantum Markov chain by a local operation on the purifying system, as stated in the following lemma. We will use this equivalence for analyzing relations among bound sets and monotones.

\blmm{sepandmark}
The following three conditions are equivalent:
\benum
\item $\rho^{AB}$ is separable.
\item There exists an extension $\vro^{ABE}$ of $\rho^{AB}$ satisfying $\vro^{ABE}\in{\ca S}_{\rm Markov}^{A:B|E}$.
\item For any purification $\ket{\varphi_\rho}^{ABE}$ of $\rho^{AB}$, there exists an operation ${\ca T}$ on $E$ such that ${\ca T}(\varphi_\rho)\in{\ca S}_{\rm Markov}^{A:B|E}$.
\ennum
\elmm

\bprf
1)$\Rightarrow$2) was proved in \cite{christandl04} (see the proof of Theorem 7 therein), and 2)$\Rightarrow$1) was proved in \cite{brandao11} (see Equality (6) therein). Thus it suffices to prove 1)$\Rightarrow$3) and 3)$\Rightarrow$2).

1)$\Rightarrow$3): Suppose that $\rho^{AB}$ is separable. By definition, there exists states $\sigma_j$, $\tau_j$ and a probability distribution $\{p_j\}_j$ such that
\alg{
\rho^{AB}=\sum_jp_j\sigma_j^{A}\otimes\tau_j^{B}.\nn
}
Thus a purification $\ket{{\tilde\varphi}_\rho}^{ABE}$ of $\rho^{AB}$ is given by
\alg{
\ket{{\tilde\varphi}_\rho}^{ABE}=\sum_j\sqrt{p_j}\ket{\phi_{\sigma_j}}^{A}\ket{\phi_{\tau_j}}^{B}\ket{j}^{E}.\nn
}
Due to Uhlmann's theorem\cite{uhlmann1976transition}, for any purification $\ket{\varphi_\rho}^{ABE}$ of $\rho^{AB}$, there exists a linear isometry ${\ca W}$ on $E$ such that $({\rm id}^{AB}\otimes{\ca W})(\proj{\varphi_\rho})=\proj{{\tilde\varphi}_\rho}$. Let $\ca D$ be the dephasing operation on $E$ with respect to the basis $\{\ket{j}\}_j$, and define ${\ca T}:={\ca D}\circ{\ca W}$. Then we have
\alg{
{\ca T}(\varphi_\rho)={\ca D}({\tilde\varphi}_\rho)=\sum_jp_j\sigma_j^{A}\otimes\tau_j^{B}\otimes\proj{j}^{E}\in{\ca S}_{\rm Markov}^{A:B|E},\nn
}
which completes the proof of 1)$\Rightarrow$3).

3)$\Rightarrow$2): Suppose that, for a purification $\ket{\varphi_\rho}^{ABE}$ of $\rho^{AB}$, there exists an operation ${\ca T}$ on $E$ such that ${\ca T}(\varphi_\rho)\in{\ca S}_{\rm Markov}^{A:B|E}$. It is straightforward to verify that ${\ca T}(\varphi_\rho)$ is an extension of $\rho^{AB}$. Hence 3) implies 2). \QED

\eprf

Due to \rLmm{sepandmark}, the bound sets ${\ca S}_J^*$ and ${\ca S}_J$ are represented as
\alg{
&{\ca S}_J^*=\{\rho\in{\ca S}_{\rm all}\:|\:\exists\vro^{ABEF},\:\exists{\ca W}:E\rightarrow E_AE_B\text{ s.t. }\nn\\
&\quad\quad\quad\varrho^{ABE}=\rho^{ABE},\;{\ca W}(\vro^{ABEF})\in{\ca S}_{\rm Markov}^{AE_A:BE_B|F}\}\laeq{koikoi}
}
and
\alg{
&{\ca S}_J=\{\rho\in{\ca S}_{\rm all}\:|\:\exists|\phi_\rho\rangle^{ABEF_AF_B},\;\exists{\ca T}\in{\mathbb L}_E\text{ s.t. }\nn\\
&\quad\phi_\rho^{ABE}=\rho^{ABE},\;{\ca T}(\phi_\rho)^{ABEF_AF_B}\in{\ca S}_{\rm Markov}^{AF_A:BF_B|E}\},\laeq{koikoikoi}
}
respectively. Note that ${\ca W}(\vro^{ABEF})$ in \req{koikoi} is an extension of ${\ca W}(\rho^{ABE})$. Hence, in the same way as  \req{idasm}-\req{idasmii}, we have
\alg{
&J_\downarrow^*(\rho)=0\text{ for any }\rho\in{\ca S}_J^*,\laeq{idasmiii}\\
&J_\downarrow(\rho)=0\text{ for any }\rho\in{\ca S}_J.\laeq{idasmiiii}
}

\subsection{Proof of Inclusion Relations}\lsec{inclusion}

In the following, we prove Relations \req{kudakero}, \req{moushikomi} and \req{kukukku}, except the strictness of their relations. Relation (\ref{eq:atatte}) and the strictness of (\ref{eq:kudakero}), (\ref{eq:moushikomi}) will be proved in Section \rsec{unitres} by construction.

\subsubsection{Relation \req{kudakero}}
Note that any state $\rho\in{\ca S}_{\rm Markov}$ is decomposed by a linear isometry $\Gamma$ from $E$ to $E_0E_LE_R$ in the form of \req{decomposability}. Denoting $E_0E_L$ and $E_R$ by $E_A$ and $E_B$, respectively, it follows that the state \req{decomposability} is separable between $AE_A$ and $BE_B$, which implies ${\ca S}_J^*\supseteq{\ca S}_{\rm Markov}$. Let $F_0$, $F_L$ and $F_R$ be quantum systsmes, and $\ket{\phi_{\sigma_j}}^{AE_LF_L}$ and $\ket{\phi_{\tau_j}}^{BE_RF_R}$ be purifications of $\sigma_j$ and $\tau_j$, respectively, for each $j$. A purification $|\phi_\rho\rangle$ of $\rho$ is given by
\alg{
&\ket{\phi_\rho}^{ABEF_0F_LF_R}:=\nn\\
&\quad\quad\quad\Gamma^\dagger\sum_j\sqrt{p_j}\ket{j}^{E_0}\ket{j}^{F_0}\ket{\phi_{\sigma_j}}^{AE_LF_L}\ket{\phi_{\tau_j}}^{BE_RF_R},\nn
}
from which, by tracing out $E$, we obtain
\alg{
{\phi_\rho}^{ABF_0F_LF_R}=\sum_jp_j\proj{j}^{F_0}\otm\phi_{\sigma_j}^{AF_L}\otm\phi_{\tau_j}^{BF_R}.\nn
}
Denoting $F_0F_L$ and $F_R$ by $F_A$ and $F_B$, respectively, the above state is separable between $AF_A$ and $BF_B$, which leads to ${\ca S}_J\supseteq{\ca S}_{\rm Markov}$.

\subsubsection{\it Relation \req{moushikomi}}
Suppose $\rho\in{\ca S}_I$. There exists an operation ${\ca T}\in{\mathbb L}_E$ such that ${\ca T}(\rho)\in{\ca S}_{\rm Markov}^{A:B|E}$. Since ${\ca T}(\rho)$ is an extension of $\rho^{AB}$, it follows from \rLmm{sepandmark} that $\rho^{AB}$ is separable between $A$ and $B$, leading to $\rho\in{\ca S}_{\rm sep}$. Hence ${\ca S}_I\subseteq{\ca S}_{\rm sep}$. Suppose $\rho\in{\ca S}_I^*$. There exists an extension $\vro^{ABEF}$ of $\rho$ such that $\vro^{ABF}\in{\ca S}_{\rm Markov}^{A:B|F}$. \rLmm{sepandmark} implies separability of $\rho^{AB}$, which yields $\rho\in{\ca S}_{\rm sep}$ and thus ${\ca S}_I^*\subseteq{\ca S}_{\rm sep}$.

\subsubsection{\it Relation \req{kukukku}}
Suppose $\rho\in{\ca S}_J$. There exists a purification $|\phi_\rho\rangle^{ABEF_AF_B}$ of $\rho$ such that $\phi_\rho^{ABF_AF_B}\in{\ca S}_{\rm sep}^{AF_A:BF_B}$. Due to \rLmm{sepandmark}, there exists an operation ${\ca T}\in{\mathbb L}_E$ such that ${\ca T}(\phi_\rho)\in{\ca S}_{\rm Markov}^{AF_A:BF_B|E}$. Noting that
\alg{
{\rm Tr}_{F_AF_B}[{\ca T}(\phi_\rho)]={\ca T}(\rho),\nn
}
we obtain
\alg{
I(AF_A:BF_B|E)_{{\ca T}(\phi_\rho)}
&\geq I(A:B|E)_{{\ca T}(\phi_\rho)}\nn\\
&=I(A:B|E)_{{\ca T}(\rho)}\nn
} 
by the data processing inequality, which implies $I(A:B|E)_{{\ca T}(\rho)}=0$ and thus ${\ca T}(\rho)\in{\ca S}_{\rm Markov}^{A:B|E}$. Hence $\rho\in{\ca S}_I$, which implies ${\ca S}_J\subseteq{\ca S}_I$. 

Suppose $\rho\in{\ca S}_J^*$. There exists a linear isometry $\ca W$ from $E$ to $E_AE_B$ such that ${\ca W}(\rho^{ABE})\in{\ca S}_{\rm sep}^{AE_A:BE_B}$. Due to \rLmm{sepandmark}, there exists an extension $\vro'^{AE_ABE_BF}\in{\ca S}_{\rm Markov}^{AE_A:BE_B|F}$ of ${\ca W}(\rho)$. It is straightforward to verify that $\vro^{ABEF}:={\ca W}^{-1}(\vro')$ is an extension of $\rho^{ABE}$. In addition, due to $\vro^{ABF}=\vro'^{ABF} $we have
\alg{
I(AE_A:BE_B|F)_{\vro'}\geq I(A:B|F)_{\vro'}=I(A:B|F)_{\vro},\nn
} 
which yields $I(A:B|F)_{\vro}$. Hence $\rho\in{\ca S}_I^*$ and thus ${\ca S}_J^*\subset{\ca S}_J$.

\QED

\hfill

\section{Duality of Monotones and Bound Sets}\lsec{duality}

Consider a tripartite quantum state $\rho$ on system $ABE$, and let $|\phi_\rho\rangle$ be an arbitrary purification thereof, on system $ABEF$. The CQMI satisfies a relation
\alg{
I(A:B|E)_{\phi_\rho}=I(A:B|F)_{\phi_\rho},\nn
}
which is often referred to as the {\it duality} of CQMI\cite{berta2015renyi}. It follows that the condition of a tripartite quantum state $\rho$ on system $ABE$ being a quantum Markov chain conditioned by $E$ is equivalent to the condition that its purification $|\phi_\rho\rangle$ on $ABEF$ is, considering the reduced state on $ABF$, a quantum Markov chain conditioned by $F$. The following proposition states that a similar type of duality holds for the monotone functions and the bound sets introduced in Sections \rsec{NMmonot} and \rsec{boundset}, respectively.

\bprp{pfeqmonot}
Let $|\phi_1\rangle^{ABEF}$ and $|\phi_2\rangle^{ABEF}$ be purifications of $\rho_1,\rho_2\in{\ca S}_{\rm all}$, respectively, and let ${\ca U}_{\rm SWAP}^{EF}$ be the swap operation of $E$ and $F$. If
\alg{
{\ca U}_{\rm SWAP}^{EF}(\proj{\phi_1})=\proj{\phi_2}\laeq{icannnn}
}
holds, we have
\alg{
&I_\downarrow(\rho_1)=I_\downarrow^*(\rho_2),\laeq{abehiro}\\
&J_\downarrow(\rho_1)=J_\downarrow^*(\rho_2).\laeq{nakama}
}
Under the same condition, we also have
\alg{
&\rho_1\in{\ca S}_I\;\Leftrightarrow\;\rho_2\in{\ca S}_I^*,\laeq{FBI}\\
&\rho_1\in{\ca S}_J\;\Leftrightarrow\;\rho_2\in{\ca S}_J^*.\laeq{CIA}
}
\eprp

\bprf

As stated in \cite{christandl04} (see Section I\!I\!I therein), for any operation $\tilde{\ca T}$ on $F$, the state ${\tilde{\ca T}}(\proj{\phi_2})^{ABEF}$ is an extension of $\rho_2^{ABE}$. Conversely, for any extension $\vro_2^{ABEF}$ of $\rho_2^{ABE}$, there exists an operation $\tilde{\ca T}$ on $F$ such that ${\tilde{\ca T}}(\proj{\phi_2})^{ABEF}=\vro_2^{ABEF}$. In the same way, for any isometry $\ca W$ from $F$ to $F_AF_B$, the state ${\ca W}(\proj{\phi_1})^{ABEF_AF_B}$ is a purification of $\rho_1^{ABE}$, and conversely, for any purification $|\phi_1'\rangle^{ABEF_AF_B}$ of $\rho_1^{ABE}$, there exists an isometry $\ca W$ from $F$ to $F_AF_B$ such that ${\ca W}(\proj{\phi_1})^{ABEF_AF_B}=\proj{\phi_1'}$.

Let $\ca T$ be an arbitrary operation on $E$, and let $\tilde{\ca T}$ be an operation on $F$ that acts in the same way as $\ca T$. It follows that
\alg{
I(A:B|E)_{{\ca T}(\rho_1)}&=I(A:B|E)_{{\ca T}(\phi_1)}=I(A:B|F)_{{\tilde{\ca T}}(\phi_2)}.\laeq{icannot}
}
Taking the infimum over all $\ca T$, we obtain
\alg{
\inf_{\ca T}I(A:B|E)_{{\ca T}(\rho_1)}&=\inf_{\tilde{\ca T}}I(A:B|F)_{{\tilde{\ca T}}(\phi_2)}\nn\\
&=\inf_{\vro_2}I(A:B|F)_{\vro_2},\nn
}
which implies \req{abehiro}. Due to \req{icannot}, the following conditions are equivalent:
\benum
\item existence of $\ca T$ satisfying $I(A:B|E)_{{\ca T}(\rho_1)}=0$,
\item existence of $\tilde{\ca T}$ satisfying $I(A:B|F)_{{\tilde{\ca T}}(\phi_2)}=0$. 
\ennum
In addition, due to the argument in the first paragraph, Condition 2) is equivalent to the existence of $\vro_2$ satisfying $I(A:B|F)_{\vro_2}=0$. Hence we obtain \req{FBI}.

To prove \req{nakama}, let $\ca W$ be an arbitrary linear isometry from $F$ to $F_AF_B$, and let $\tilde{\ca W}$ be a linear isometry from $E$ to $E_AE_B$ that acts in the same way as $\ca W$. We have
\alg{
&I(AF_A:BF_B|E)_{({\ca W}\otimes{\ca T})(\phi_1)}\nn\\
&=I(AE_A:BE_B|F)_{({\tilde{\ca W}}\otm{\tilde{\ca T}})(\phi_2)}.\nn
}
Taking the infimum over all $\ca T$, we obtain
\alg{
&\inf_{\ca T}I(AF_A:BF_B|E)_{({\ca W}\otimes{\ca T})(\phi_1)}\nn\\
&=\inf_{\tilde{\ca T}}I(AE_A:BE_B|F)_{({\tilde{\ca W}}\otm{\tilde{\ca T}})(\phi_2)}\nn\\
&=\inf_{\vro_2}I(AE_A:BE_B|F)_{{\tilde{\ca W}}(\vro_2)},\nn
}
where the infimum in the last line is taken over all extensions $\vro_2$ on $ABF_AF_B$ of $\rho$.  By further taking the infimum over all $\ca W$, we arrive at
\alg{
&\inf_{\phi_1'}\inf_{\ca T}I(AF_A:BF_B|E)_{{\ca T}(\phi_1')}\nn\\
&=\inf_{\ca W}\inf_{\ca T}I(AF_A:BF_B|E)_{({\ca W}\otimes{\ca T})(\phi_1)}\nn\\
&=\inf_{\tilde{\ca W}}\inf_{\vro_2}I(AE_A:BE_B|F)_{{\tilde{\ca W}}(\vro_2)},\nn
}
where the infimum in the first line is taken over all purifications $\phi_1'$ of $\rho_1$.  Thus we obtain \req{nakama}. It is straightforward to verify that the following conditions are equivalent due to \req{icannnn}:
\benum
\item existence of ${\ca W}$ such that ${\rm Tr}_{E}[{\ca W}(\proj{\phi_1})^{ABEF_AF_B}]$ is separable between $AF_A$ and $BF_B$,
\item existence of $\tilde{\ca W}$ such that ${\rm Tr}_{F}[{\tilde{\ca W}}(\proj{\phi_2})^{ABE_AE_BF}]$ is separable between $AE_A$ and $BE_B$.
\ennum
Noting that ${\ca W}(\proj{\phi_1})$ is a purification of $\rho_1$, and that we have 
\alg{
{\tilde{\ca W}}({\rho_2})^{ABE_AE_B}={\rm Tr}_{F}[{\tilde{\ca W}}(\proj{\phi_2})^{ABE_AE_BF}],\nn
}
we obtain \req{CIA}.\QED
\eprf

\hfill

\section{Examples}\label{sec:unitres}

\begin{table}[t]
\renewcommand{\arraystretch}{1.6}
  \begin{center}
    \begin{tabular}{|c|cccccc|c|} \hline
                             & $I_M$ & $I_\downarrow$ & $I_\downarrow^*$ & $I_{sq}$ & $J_\downarrow$ & $J_\downarrow^*$ &  bound set\\ \hline 
    $\Phi_{\rm I}$   & $2$     & $2$                    &  $2$                       & $2$        & $2$ & $2$       & ${\ca S}_{\rm all}\backslash{\ca S}_{\rm sep}$ \\
    $\Phi_{\rm I\!I}$ & $2$     & $1$                    &  $1$                       & $0$        & $2$ & $2$       & ${\ca S}_{\rm sep}\backslash{\ca S}_I\cup{\ca S}_I^*$ \\
    $\Phi_{\rm I\!I\!I}$ & $2$     & $0$                    &  $0$                       & $0$        & $0$ & $0$      & ${\ca S}_J\cap{\ca S}_J^*\backslash{\ca S}_{\rm Markov}$ \\
   $\Psi_{\rm I}^*$ & $1$     & $1$                    &  $0$                       & $0$        & $2$ & $0$        & ${\ca S}_J^*\backslash{\ca S}_I$ \\
$\Psi_{\rm I}$ & $1$     & $0$                    &  $1$                       & $0$        & $0$ & $2$        & ${\ca S}_J\backslash{\ca S}_I^*$ \\
 $\Psi_{\rm I\!I}^*$ & $1$     & $0$                    &  $0$                       & $0$         & $0$ & $0$       & ${\ca S}_J\cap{\ca S}_J^*\backslash{\ca S}_{\rm Markov}$ \\
$\Psi_{\rm I\!I}$ & $1$     & $0$                    &  $0$                       & $0$        & $0$ & $0$      & ${\ca S}_J\cap{\ca S}_J^*\backslash{\ca S}_{\rm Markov}$ \\  \hline
    \end{tabular}
  \end{center}
  \caption{values of non-Markovianity monotones}
  \label{tb:values}
\end{table}

In this section, we consider examples of states on a tripartite system composed of qubits, and analyze convertibility among those states under $\Omega$. For each state, we compute values of the non-Markovianity monotones introduced in \rSec{NMmonot}, as well as identify bound sets that the state belongs to. Examples presented here completes the proof of Relations \req{kudakero}-\req{kukukku}. We also provide examples to show that {\it irreversible} operations by Eve may generate and increase non-Markovianity of quantum states.

\subsection{Convertibility among The States}

Define pure states $|\Phi_{pq}\rangle\:(p,q=0,1)$ of a system composed of two qubits by
\alg{
|\Phi_{pq}\rangle:=\frac{1}{\sqrt{2}}(\sigma_x^p\sigma_z^q\otimes I)\left(\ket{00}+\ket{11}\right),\nn
}
where $\sigma_x$ and $\sigma_z$ are Pauli operators defined by $\sigma_x:=|0\rangle\!\langle1|+|1\rangle\!\langle0|$ and $\sigma_z:=|0\rangle\!\langle0|-|1\rangle\!\langle1|$. Define a state $\Phi_{\rm I}\in{\ca S}_{\rm all}$ by
\alg{
\ket{\Phi_{\rm I}}^{ABE}:=|\Phi_{00}\rangle^{AB}\ket{0}^E.\nn
}
Consider the following protocol that belongs to $\Omega$:
\begin{itemize}
 \setlength{\leftskip}{0.3cm}
\item[(P1)] Alice flips a fair coin and broadcasts the result ($c=0,1$) to Bob and Eve. Bob discards the message communicated from Alice, while Eve records it on her quantum register. Alice performs $\sigma_z$ on her qubit if $c=1$, whereas she does not apply any operation if $c=0$. Finally Alice erases the memory in which the result of coin flip has been recorded.
\end{itemize}
It is straightforward to verify that the state $\Phi_{\rm I}$ is transformed by the above protocol to the state
\alg{
\Phi_{\rm I\!I}^{ABE}:=&\frac{1}{2}\:|\Phi_{00}\rangle\!\langle\Phi_{00}|^{AB}\otimes\proj{0}^E\nn\\
&\quad+\frac{1}{2}\:|\Phi_{01}\rangle\!\langle\Phi_{01}|^{AB}\otimes\proj{1}^E.\laeq{hashiri}
}
Let (P1') be the same protocol as (P1), except that Alice applies $\sigma_z$ instead of $\sigma_x$ if $c=1$. This protocol transforms $\Phi_{\rm I\!I}$ to
\alg{
\Phi_{\rm I\!I\!I}^{ABE}:=\:\frac{1}{4}\sum_{p,q=0,1}|\Phi_{pq}\rangle\!\langle\Phi_{pq}|^{AB}\otimes\proj{p,q}^E.\laeq{surr}
}

Define a state $\Psi_{\rm I}^*$ by
\alg{
\Psi_{\rm I}^{*ABE}:=\frac{1}{2}\left(\proj{00}+\proj{11}\right)^{AB}\otimes\proj{0}^E.\laeq{evol}
}
Let ${\ca D}$ be the dephasing operation on $A$ with respect to the basis $\{\ket{0},\ket{1}\}$, represented by ${\ca D}(\cdot)=\proj{0}(\cdot)\proj{0}+\proj{1}(\cdot)\proj{1}$, and $\ca V$ be a reversible operation on $E$ defined by 
\alg{
{\ca H}_{{\rm In}({\ca V})}={\mathbb C},\;{\ca H}_{{\rm Out}({\ca V})}={\mathbb C}^2,\;{\ca V}(\proj{0})=\pi_2.\nn
}
We have ${\ca D}(\Phi_{\rm I\!I})={\ca V}(\Psi_{\rm I}^*)$, hence $\Phi_{\rm I\!I}$ is convertible to $\Psi_{\rm I}^*$. The protocol (P1') transforms $\Psi_{\rm I}^*$ to
\alg{
\Psi_{\rm I\!I}^{*ABE}:=&\frac{1}{4}\left(\proj{00}+\proj{11}\right)^{AB}\otimes\proj{0}^E\nn\\
&\quad+\frac{1}{4}\left(\proj{01}+\proj{10}\right)^{AB}\otimes\proj{1}^E.\nn
}

\begin{figure}[t]
\begin{center}
\includegraphics[bb={0 0 356 407}, scale=0.4]{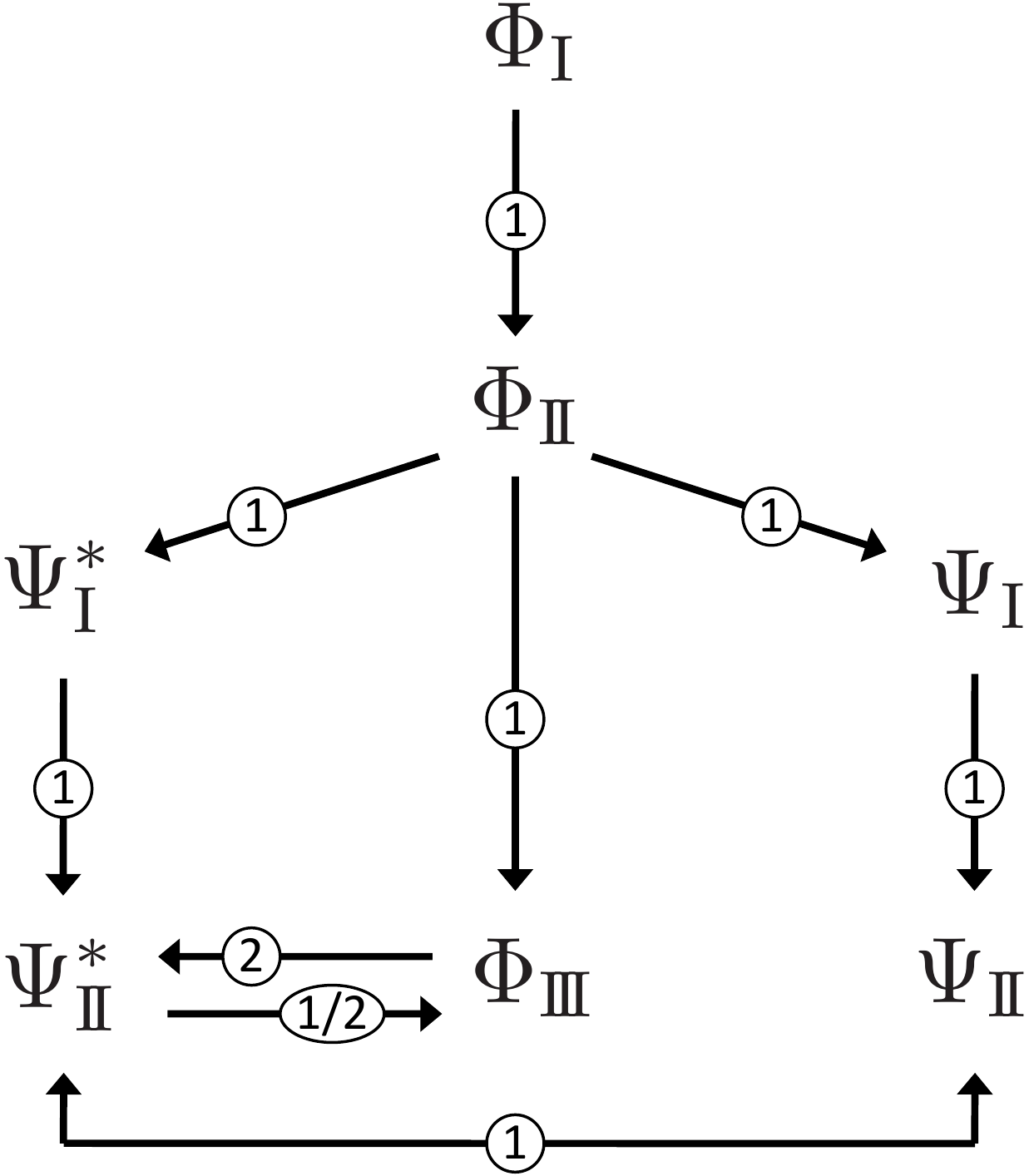}
\end{center}
\caption{Convertibility between the states is depicted. The number on each arrow represents the optimal rate of number of copies at which the initial state is convertible to the final state under $\Omega$. The optimality of those ratios follows from the values of non-Markovianity monotones for each state (see Table \ref{tb:values}).}
\label{fig:convertibility}
\end{figure}

Let $\Psi_{\rm I}$ be the GHZ state\cite{greenberger1989going} defined by
\alg{
|\Psi_{\rm I}\rangle^{ABE}:=\frac{1}{\sqrt{2}}\left(\ket{000}+\ket{111}\right)^{ABE}.\laeq{mouichi}
}
Suppose that the three parties initially share $\Phi_{\rm I\!I}$. Alice prepares a qubit system $\tilde E$ in the state $|0\rangle$, performs the CNOT gate
\alg{
U_{\rm CNOT}^{A\tilde E}:=\proj{0}^A\otimes I^{\tilde E}+\proj{1}^A\otimes \sigma_x^{\tilde E}\nn
}
and sends $\tilde E$ to Eve. Eve then performs the CZ gate
\alg{
U_{\rm CZ}^{E\tilde E}:=\proj{0}^E\otimes I^{\tilde E}+\proj{1}^E\otimes \sigma_z^{\tilde E}.\nn
}
Noting that we have
\alg{
U_{\rm CZ}^{E\tilde E}U_{\rm CNOT}^{A\tilde E}|\Phi_{0p}\rangle^{AB}|0\rangle^{\tilde E}|p\rangle^{E}=|\Psi_{\rm I}\rangle^{AB{\tilde E}}|p\rangle^{E}\nn
}
for $p=0,1$, this protocol transforms $\Phi_{\rm I\!I}$ to the state
\alg{
\Psi_{\rm I}^{AB{\tilde E}}\otimes\pi_2^{E}={\ca V}'(\Psi_{\rm I}),\nn
}
where ${\ca V}'$ is a reversible operation on $E$ defined by
\alg{
{\ca H}_{{\rm In}({\ca V})}={\mathbb C}^2,\;{\ca H}_{{\rm Out}({\ca V})}=({\mathbb C}^2)^{\otimes 2},\;{\ca V}(\rho)=\rho\otm\pi_2\nn
}
and $\pi_2=(\proj{0}+\proj{1})/2$. Hence $\Phi_{\rm I\!I}$ is convertible to $\Psi_{\rm I}$.

Suppose that the three parties initially share $\Psi_{\rm I}$. They apply the protocol (P1'), and then Eve performs the Hadamard operation $H$ defined by
\alg{
H:=(\outpro{0}{0}+\outpro{0}{1}+\outpro{1}{0}-\outpro{1}{1})/\sqrt{2}\nn
} 
on her qubit. Defining 
\alg{
|\psi_p\rangle^{ABE}:=((\sigma_x^p)^A\otimes I^{B}\otimes H^E)|\Psi_{\rm I}\rangle\;(p=0,1),\laeq{watatame}
}
 the obtained state is expressed as
\alg{
\Psi_{\rm I\!I}^{ABE}:=&\:\frac{1}{2}(\proj{\psi_0}\otimes\proj{0}+\proj{\psi_1}\otimes\proj{1})\label{eq:schwartz}\\
=&\:\frac{1}{4}\sum_{p,q,q'=0,1}|\Phi_{pq}\rangle\!\langle\Phi_{pq'}|^{AB}\otimes\ket{p,q}\!\bra{p,q'}^E,\nn
}
where the second line follows from
\alg{
|\psi_p\rangle^{ABE}=\frac{1}{\sqrt{2}}(|\Phi_{p0}\rangle^{AB}\ket{0}^E+|\Phi_{p1}\rangle^{AB}\ket{1}^E).\nn
}

We denote $\rho\sim\sigma$ if a state $\rho\in{\ca S}_{\rm all}$ is convertible to $\sigma\in{\ca S}_{\rm all}$ and vice versa. As we prove in Appendix \ref{app:prforaora}, it holds that
\alg{
\Psi_{\rm I\!I}\sim\Psi_{\rm I\!I}^*,\quad\Phi_{\rm I\!I\!I}\sim\Psi_{\rm I\!I}^{*\otimes 2}.\label{eq:euston}
}
The convertibility among the states presented in this section is depicted in \rFig{convertibility}.

We will prove in Appendix \ref{app:increl} that we have
\alg{
&\Phi_{\rm I}\in{\ca S}_{\rm all}\backslash{\ca S}_{\rm sep},\label{eq:phiinsep}\\
&\Phi_{\rm I\!I}\in{\ca S}_{\rm sep}\backslash{\ca S}_I\cup{\ca S}_I^*,\label{eq:psiinzi}\\
&\Psi_{\rm I}^*\in{\ca S}_J^*\backslash{\ca S}_I,\label{eq:psiinzii}\\
&\Psi_{\rm I}\in{\ca S}_J\backslash{\ca S}_I^*,\label{eq:psiinziii}\\
&\Phi_{\rm I\!I\!I},\Psi_{\rm I\!I}^*,\Psi_{\rm I\!I}\in{\ca S}_J\cap{\ca S}_J^*\backslash{\ca S}_{\rm Markov}\label{eq:roki}
}
The above relations provide a proof for Relation (\ref{eq:atatte}) and the strictness of (\ref{eq:kudakero}) and (\ref{eq:moushikomi}). The values of non-Markovianity monotones for the above states are listed in Table \ref{tb:values}.

\subsection{Irreversible Operations}

Suppose Alice flips a fair coin, record the result on her quantum register, and broadcasts it to Bob and Eve, who also record it on their quantum register to obtain the state
\alg{
\Psi_{\rm I}'^{ABE}:=\frac{1}{2}\left(\proj{000}+\proj{111}\right)^{ABE}.\nn
}
The state $\Psi_{\rm I}^*$ in \req{evol} is obtained from this state by Eve performing an irreversible operation $\ca E$ defined by ${\ca E}(\tau)=\proj{0}$ for all input state $\tau$.

The Hadamard operation by Eve transforms the state $\Psi_{\rm I}$ defined by \req{mouichi} to the state
\alg{
|\Psi_{\rm I}'\rangle^{ABE}
&:=(I^{AB}\otimes H^E)|\Psi_{\rm I}\rangle^{ABE}\nn\\
&=\frac{1}{\sqrt{2}}\left(\ket{00}^{AB}\ket{+}^E+\ket{11}^{AB}\ket{-}^E\right)\nn\\
&=\frac{1}{\sqrt{2}}\left(\ket{\Phi_{00}}^{AB}\ket{0}^E+\ket{\Phi_{01}}^{AB}\ket{1}^E\right).\nn
}
By dephasing operation on $E$ with respect to the basis $\{\ket{0},\ket{1}\}$, this state is transformed to $\Phi_{\rm I\!I}$ defined by \req{hashiri}.

Let $E_A$ and $E_B$ be quantum registers possessed by Eve, and define a pure state $|{\bf\Phi}\rangle^{ABE_AE_B}:=|\Phi_{00}\rangle^{AE_A}|\Phi_{00}\rangle^{BE_B}$. This state is prepared from scratch by quantum communication from Alice and Bob to Eve. A Schimidt decomposition of this state between $AB$ and $E_AE_B$ is given by
\alg{
|{\bf\Phi}\rangle^{ABE_AE_B}=\frac{1}{2}\sum_{p,q=0,1}|\Phi_{pq}\rangle^{AB}|\Phi_{pq}\rangle^{E_AE_B}.\nn
}
By an irreversible operation $\ca E$ on $E_AE_B$ defined by 
\alg{
{\ca E}(\cdot)=\sum_{p,q=0,1}\outpro{p,q}{\Phi_{pq}}(\cdot)\outpro{\Phi_{pq}}{p,q},\nn
}
this state is transformed to $\Phi_{\rm I\!I\!I}$ defined by \req{surr}.

\hfill

\section{Non-Markovianity Dilution}\lsec{disdil}

In this section, we introduce a task of {\it non-Markovianity dilution}, and define the {\it non-Markovianity cost} of a state as the optimal achievable rate in non-Markovianity dilution. We prove that the non-Markovianity cost for pure states is bounded from above the entanglement of purification of its bipartite reduced state.

\subsection{Definitions}

Let us consider asymptotic convertibility of states under operations in $\Omega$. We generalize the definition of state convertibility (\rdfn{zconvertibility}) to that of $\epsilon$-convertibility as follows:
\bdfn{econvertibility}
A state $\rho_1$ is {\it $\epsilon$-convertible to $\rho_2$ under $\Omega$} if there exists an operation ${\ca E}\in\Omega$ and a reversible operation $\ca V$ on $E$ such that
\alg{
\left\|{\ca E}(\rho_1^{ABE})-{\ca V}(\rho_2^{ABE})\right\|_1\leq\epsilon,\nn
}
where $\norm{\cdot}_1$ is the trace norm defined by $\norm{A}_1={\rm Tr}|A|$ for an operator $A$.
\edfn

In analogy to entanglement dilution, we formulate non-Markovianity dilution as a task in which copies of the maximally non-Markovian state is transformed by an operation in $\Omega$ to copies of a state we are concerning. We require that the error vanishes in the asymptotic limit of infinite copies. The non-Markovianity cost of a state is defined as the minimum number of copies of the maximally non-Markovian state required in non-Markovianity dilution. Rigorous definitions of non-Markovianity dilution and the non-Markovianity cost are given as follows:

\bdfn{nMcost}
A rate $R$ is {\it achievable in non-Markovianity dilution of a state $\rho\in{\ca S}_{\rm all}$} if, for any $\epsilon>0$ and sufficiently large $n$, the state $\Phi_{\rm I}^{\otimes nR}$ is $\epsilon$-convertible to $(\rho^{ABE})^{\otimes n}$ under $\Omega$. The {\it non-Markovianity cost} of a state $\rho$, which we denote by $M_C(\rho)$, is defined as the infimum of achievable $R$ in non-Markovianity dilution of $\rho$.
\edfn

\subsection{Non-Markovianity Cost of Pure States}

The {\it entanglement of purification}\cite{terhal2002entanglement} of a bipartite quantum state $\tau$ on system $AB$ is defined as
\alg{
E_P(\tau):=&\inf_{\phi_\tau}S(AE_A)_{\phi_\tau}\nn\\
=&\inf_{\ca W}S(AE_A)_{{\ca W}(\phi_\tau^*)},\nn
}
where the infimum in the first line is taken over all purifications $\phi_\tau$ of $\tau$ on $AE_ABE_B$, and one in the second line over all linear isometry from $E$ to $E_AE_B$ with a fixed purification $\phi_\tau^*$ of $\tau$ on $ABE$. It is straightforward to verify that, for any tripartite pure state $|\psi\rangle$ on system $ABE$, we have
\alg{
J_\downarrow^*(\psi)=2E_P(\psi^{AB}).\nn
} 
The {\it regularized entanglement of purification} of a bipartite state $\tau$ is defined as
\alg{
E_P^\infty(\tau):=\lim_{n\rightarrow\infty}\frac{1}{n}E_P(\tau^{\otimes n}).\nn
}
The following proposition states that the non-Markovianity cost of a pure state is bounded from above by the regularized entanglement of purification.

\bthm{nMcEP}
For all tripartite pure state $|\psi\rangle$ on $ABE$, we have
\alg{
M_C(\psi)\leq E_P^\infty(\psi^{AB}).\laeq{oboeteru}
}
\ethm

\bprf
It was proved in \cite{terhal2002entanglement} that the regularized entanglement of purification is equal to the asymptotic cost of entanglement that is required for generating copies of a bipartite quantum state only by local operations (without communication). In particular, it was proved in \cite{terhal2002entanglement} that for any $R>E_P^\infty(\psi^{AB})$, $\epsilon>0$ and sufficiently large $n$, there exists local operations ${\ca E}_1$ by Alice and ${\ca E}_2$ by Bob such that
\alg{
\left\|({\ca E}_1^A\otm{\ca E}_2^B)(\Phi_2^{\otimes nR})-(\psi^{AB})^{\otm n}\right\|_1\leq\epsilon.\laeq{evoevo}
}
Let $A_0$ and $B_0$ be ancillary system possessed by Alice and Bob, respectively, and let ${\ca U}_1:A\rightarrow AA_0$ and ${\ca U}_2:B\rarrow BB_0$ be linear isometries such that Stinespring dilations of ${\ca E}_1$ and ${\ca E}_2$ are given by ${\ca E}_1={\rm Tr}_{A_0}\circ{\ca U}_1$ and ${\ca E}_2={\rm Tr}_{B_0}\circ{\ca U}_2$, respectively. Suppose Alice and Bob initially share $nR$ copies of $\Phi_2$ and consider the following protocol that is an element of $\Omega$:
\benum
\item Alice and Bob locally performs ${\ca U}_1$ and ${\ca U}_2$, respectively.
\item Alice and Bob sends $A_0$ and $B_0$, respectively, to Eve.
\ennum
It follows from \req{evoevo} that 
\alg{
\left\|{\rm Tr}_{A_0B_0}[({\ca U}_1\otm{\ca U}_2)(\Phi_2^{\otimes nR})]-(\psi^{AB})^{\otm n}\right\|_1\leq\epsilon.\nn
}
Therefore, due to Uhlmann's theorem (\!\cite{uhlmann1976transition}, see also Lemma 2.2 in \cite{deve08}), there exists a linear isometry ${\tilde{\ca U}}$ from $A_0B_0$ to $E$ such that
\alg{
\left\|({\tilde{\ca U}}\circ({\ca U}_1\otm{\ca U}_2))(\Phi_2^{\otimes nR})-(\psi^{ABE})^{\otm n}\right\|_1\leq2\sqrt{\epsilon}.\nn
}
Hence $\Phi_2^{\otimes nR}$ is $\epsilon$-convertible to $(\psi^{ABE})^{\otm n}$ under $\Omega$, which yields $R\leq M_C(\psi)$. Since this relation holds for any $R>E_P^\infty(\psi^{AB})$, we obtain $M_C(\psi)\leq E_P^\infty(\psi)$. \QED
\eprf

It should be noted that Equality in \req{oboeteru} holds if $J_\downarrow^*$ satisfies a property called {\it asymptotic continuity}.

\hfill

\section{Classical Resource Theory}\label{sec:classical}

In this section, we consider a {\it classical} resource theory of non-Markovianity. We present settings of the problem and briefly review the main results without providing proofs, since the proofs proceeds almost in parallel to those for quantum resource theory which are provided in the previous sections and Appendices.

\subsection{Settings}

\begin{figure}[t]
\begin{center}
\includegraphics[bb={0 0 564 375}, scale=0.4]{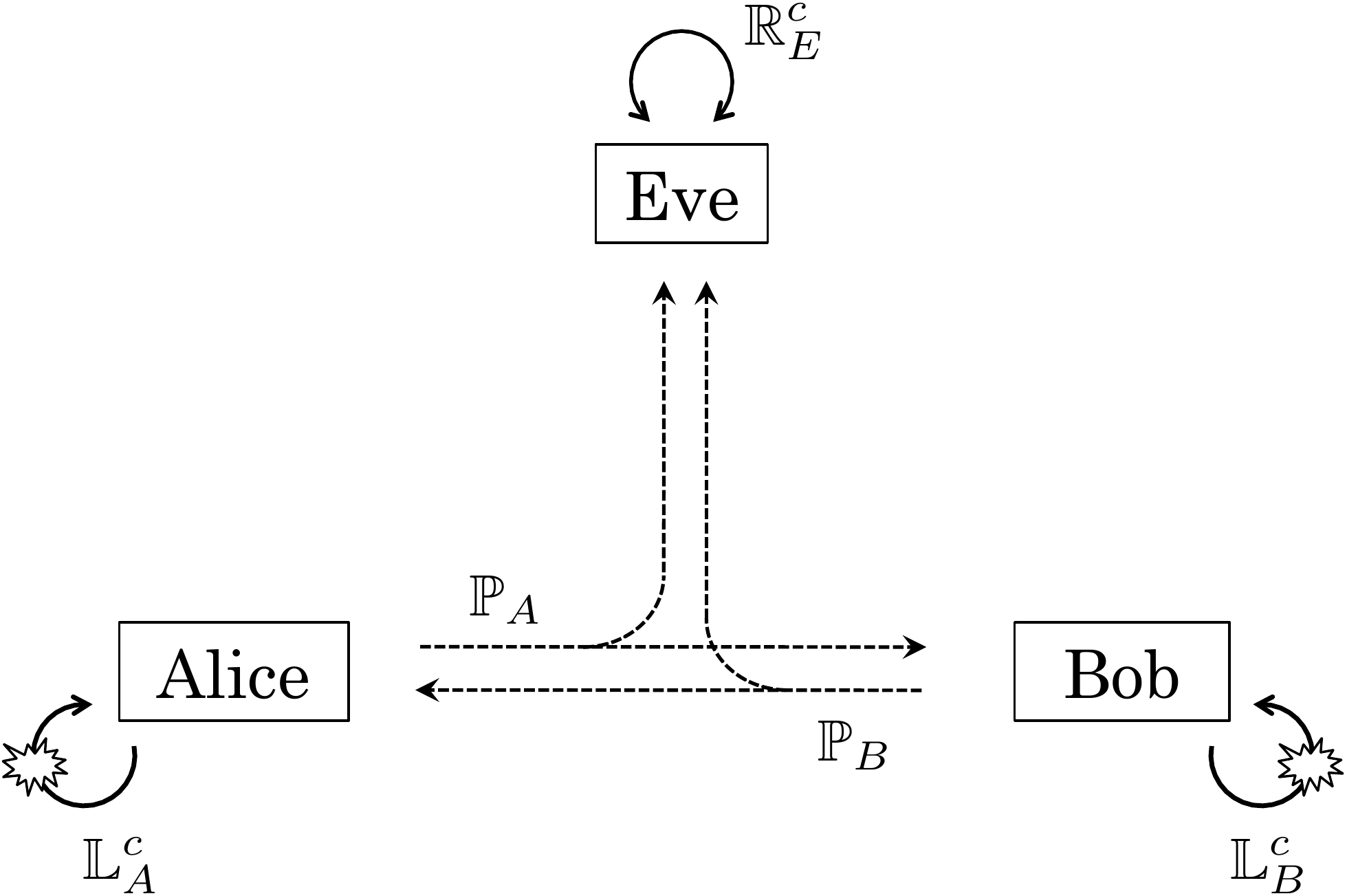}
\end{center}
\caption{The classes of operations that comprises $\Omega^c$ are depicted.}
\label{fig:classicalfreeOP}
\end{figure}

Suppose Alice, Bob and Eve have access to random variables $X$, $Y$ and $Z$, respectively. A state of those random variables is specified by finite sets $\ca X$, $\ca Y$ and $\ca Z$, in which the variables takes values, and a probability distribution $P$ on ${\ca X}\times{\ca Y}\times{\ca Z}$. We denote the set of all states by ${\ca S}_{\rm all}^c$. We consider a scenario in which the three parties are only allowed to perform the following operations and their compositions:
\alg{
{\mathbb L}_{A}^c&\text{: local processing by Alice}\nn\\
{\mathbb L}_{B}^c&\text{: local processing by Bob}\nn\\
{\mathbb R}_E^c&\text{: local reversible processing by Eve}\nn\\
{\mathbb P}_A&\text{: broadcasting of classical messages by Alice}\nn\\
{\mathbb P}_B&\text{: broadcasting of classical messages by Bob}\nn
}
By ``local processing'', we refer to an operation represented by a stochastic map from a set of alphabets to another. For example, a local processing ${\ca E}$ by Alice is represented by a conditional probability distribution $\{\{e(x'|x)\}_{x'\in{\ca X}'}\}_{x\in{\ca X}}$. A local processing ${\ca V}$ represented by $\{\{r(z'|z)\}_{z'\in{\ca Z}'}\}_{z\in{\ca Z}}$ is said to be {\it reversible} if there exists another stochastic map ${\ca V}^*$ represented by $\{\{r^*(z|z')\}_{z\in{\ca Z}}\}_{z'\in{\ca Z}'}$ such that ${\ca V}^*\circ{\ca V}={\rm id}_{\ca Z}$, or equivalently,
\alg{
\sum_{z'\in{\ca Z}'}r^*(z_2|z')r(z'|z_1)=\delta_{z_1,z_2},\quad\forall z_1,z_2\in{\ca Z}.\nn
}
We denote the set of operations that can be represented as a composition of the above operations by $\Omega^c$ (see \rFig{classicalfreeOP}). Convertibility between states is defined analogously to the quantum case as follows.


\bdfn{zconvertibilityC}
{\it A state $P_1$ is convertible to $P_2$ under $\Omega^c$} if there exists an operation ${\ca E}\in\Omega^c$ and a reversible operation ${\ca V}\in{\mathbb R}_E^c$ such that ${\ca E}(P_1)={\ca V}(P_2)$.
\edfn

\subsection{Free States, Bound Sets and Monotones}

A state $Q$ is said to be a classical Markov chain if the probability is decomposed in the form of $Q(x,y,z)=Q(x|z)Q(y|z)Q(z)$, or equivalently, if it satisfies $I(X:Y|Z)_Q=0$. We denote the set of classical Markov chains by ${\ca S}_{\rm Markov}^c$.

\bprp{freestateC}
A state $Q$ is a free state under $\Omega^c$ if and only if $Q\in{\ca S}_{\rm Markov}^c$.
\eprp
Analogously to ${\ca S}_I$ defined by \req{midomi}, the following set is a bound set under $\Omega^c$:
\alg{
{\ca S}_I^c:=\{\rho\in{\ca S}_{\rm all}^c\:|\:\exists{\ca T}\in{\mathbb L}_E^c\text{ s.t. }{\ca T}(\rho)\in{\ca S}_{\rm Markov}^c\}.\nn
}
It is straightforward to verify that we have
\alg{
{\ca S}_{\rm Markov}^c\subsetneq{\ca S}_I^c\subsetneq{\ca S}_{\rm all}^c.\laeq{CEO}
}

The conditional (classical) mutual information
\alg{
I_M^c(P):=I(X:Y|Z)_P\nn
}
and the {\it intrinsic information} \cite{gisin2000linking,maurer1999unconditionally,renner2003new}, defined as
\alg{
I_{\downarrow}^c(P):=\inf_{\ca T}I(X:Y|Z)_{{\ca T}(P)},\nn
}
are monotonically non-increasing under $\Omega^c$. Here, the infimum is taken over all stochastic map ${\ca T}$ on random variable $Z$.

\subsection{Examples}

Consider states $P_{\rm I},P_{\rm I\!I}\in{\ca S}_{\rm all}^c$ defined by probability distributions 
\alg{
p_{\rm I}(x,y,z)&=\frac{1}{2}\delta_{x,y}\delta_{z,0},\nn\\
p_{\rm I\!I}(x,y,z)&=\frac{1}{4}\delta_{x\oplus y,z},\nn
}
respectively, for $x,y,z\in\{0,1\}$. Here, $\oplus$ denotes summation modulo $2$. The states satisfy
\alg{
I_M^c(P_{\rm I})=I_M^c(P_{\rm I\!I})=I_{\downarrow}^c(P_{\rm I})=1,\quad I_{\downarrow}^c(P_{\rm I\!I})=0\nn
}
and
\alg{
P_{\rm I}\in{\ca S}_{\rm all}^c\backslash{\ca S}_I^c,\quad P_{\rm I\!I}\in{\ca S}_I^c\backslash{\ca S}_{\rm Markov}^c,\nn
}
the latter of which implies the strictness of Relations \req{CEO}. 

\subsection{Maximally non-Markovian State}

Consider a state $P_{{\rm I},d}\in{\ca S}_{\rm all}^c$ defined by a probability distribution 
\alg{
p_{{\rm I},d}(x,y,z)&=\frac{1}{d}\delta_{x,y}\delta_{z,0}\nn
}
for $x,y,z\in\{1,\cdots,d\}$. The above state is regarded as a {\it $d$-dimensional maximally non-Markovian state}, since any state $P\in{\ca S}_{\rm all}^c$ is, if ${\ca X}={\ca Y}=\{1,\cdots,d\}$,  generated from $P_{{\rm I},d}$ by an operation in $\Omega^c$. To verify this, note that the state $P_{{\rm I},d}$ provides Alice and Bob a shared random variable $K$ which obeys the uniform distribution on $\{1,\cdots,d\}$ and is decoupled from Eve. Consider the following protocol which is an element of $\Omega^c$:
\benum
\item Alice locally generates random variables $X$, $Y$ and $Z$ according to a probability distribution $\{p(x,y,z)\}_{x,y,z}$ which represents state $P$.
\item Alice computes $M=Y\oplus K$.
\item Alice broadcasts $Z$ and $M$ to Bob and Eve.
\item Bob computes $Y=M\ominus K$, where $\ominus$ denotes subtraction modulo $d$. 
\item Bob discards $Z$.
\ennum
Since the random variable  $M$ is decoupled from $X$, $Y$ and $Z$, the state $P$ is obtained in this manner. 

\hfill

\section{Conclusion} \label{sec:discussion}

In this paper, we have introduced a class of quantum operations performed by three distant parties, and analyzed an operational resource theory (ORT) induced by it. We have proved that a tripartite quantum state is a free state if and only if it is a quantum Markov chain. We introduced monotone functions and bound sets that have a clear correspondence to each other. We also formulated a task of non-Markovianity dilution, and proved that the optimal rate for the task is asymptotically given by the entanglement of purification in the case of pure states. 


The result presented in this paper is a first step toward an operational resource theory of non-Markovianity. We conclude this paper by listing problems that are left as a future work:
\bitem
\item {\it single-shot convertibility}: What is the condition under which a single copy of a tripartite pure state is convertible to another under $\Omega$?
\item {\it non-Markovianity distillation}: What is a definition of non-Markovianity distillation, and what is the optimal rate?
\item {\it bound sets and monotones}: Are there any bound set and monotone function other than those presented in \rSec{boundset} and \rsec{NMmonot}?
\item {\it monotonicity of the relative entropy of recovery}: Is the relative entropy of recovery a monotone under $\Omega$?
\item {\it strictness of inclusion relations}: Are inclusion relations \req{kukukku} strict? If so, what states belong to the relative component?
\item {\it properties of non-Markovianity monotones}: Do the non-Markovianity monotones satisfy properties such as additivity, convexity (concavity) and asymptotic continuity? Note that asymptotic continuity of $J_\downarrow^*$ implies the equality in \req{oboeteru} for all pure states.
\item {\it lockability of non-Markovianity monotones}: It has been known that the squashed entanglement and the entanglement of purification are {\it lockable}\cite{christandl2005uncertainty}. Are $J_\downarrow$ and $J_\downarrow^*$ lockable as well?
\entem

\hfill

\section*{Acknowledgment}

The author thanks Francesco Buscemi for helpful discussions, and Mark Wilde for useful comments on an earlier version of this manuscript.

\hfill

\bibliographystyle{IEEEtran}
\bibliography{bibbib.bib}

\hfill

\appendices

\section{Description of Free Operations}

In this appendix, we present a mathematical description of ${\mathbb P}_A$ and ${\mathbb R}_E$, which will be used in \rApp{monotII}.

\subsection{Description of ${\mathbb P}_A$ and ${\mathbb P}_B$}

Recall that the state before and after broadcasting of classical message by Alice are represented by density operators \req{rhohati} and \req{rhohatf}, respectively.
Let ${\varrho}_i^{C_AABEF}$ be an extension of $\rho_i$ and define
\alg{
{\varrho}_f&:=\sum_mp_m\proj{m}^{C_{\!B}}\otimes\proj{m}^{C_{\!E}}\nn\\
&\quad\quad\quad\quad\quad\quad\otimes\proj{m}^{C_{\!F}}\otimes{\varrho}_m^{ABEF},\label{eq:rhohatff}
} 
where
\alg{
{\varrho}_m^{ABEF}:=p_m^{-1}\bra{m}^{C_A}{\varrho}_i^{C_AABEF}\ket{m}^{C_A}.\nn
}
It is straightforward to verify that ${\varrho}_f$ is an extension of $\rho_f$. Let $|\phi_{\rho_m}\rangle^{ABEF}$ be a purification of $\rho_m^{ABE}$ for each $m$, and let $C_F$ be a quantum system. Purifications of $\rho_i$ and $\rho_f$ are given by  
\alg{
&|\phi_{\rho_i}^*\rangle:=\sum_m\sqrt{p_m}|m\rangle^{C_A}|m\rangle^{C_{F}}|\phi_{\rho_m}\rangle^{ABEF},\laeq{watashinara}\\
&|\phi_{\rho_f}^*\rangle:=\sum_m\sqrt{p_m}|m\rangle^{C_B}|m\rangle^{C_E}|m\rangle^{C_{F}}|\phi_{\rho_m}\rangle^{ABEF},\laeq{anatanara}
}
respectively.

\subsection{Description of Reversible Operations}

The following lemma states that a reversible operation takes a simple form in terms of the Stinespring dilation.

\blmm{SSrev}
Let $\ca V$ be a reversible operation on a system $S$, and let ${\ca V}^*$ be an operation such that ${\ca V}^*\circ{\ca V}={\rm id}|_{{\ca H}_{{\rm In}({\ca V})}}$. Let $R$ be an ancillary system, and $W$ be a linear isometry from ${\ca H}_{{\rm Out}({\ca V})}$ to ${\ca H}_{{\rm Out}({\ca V})}\otimes{\ca H}^R$ such that a Stinespring dilation of ${\ca V}^*$ is given by ${\ca V}^*(\cdot)={\rm Tr}_{R}[W(\cdot)W^\dagger]$. There exists a state $\sigma_0\in{\ca S}({\ca H}^R)$ such that for any $\tau\in{\ca S}({\ca H}_{{\rm In}({\ca V})})$ we have
\alg{
W({\ca V}(\tau))W^\dagger=\tau\otimes\sigma_0^R.\laeq{maria}
}
\elmm

\bprf
Let $S'$ be a system represented by a Hilbert space ${\ca H}'$ that has the same dimension as ${\ca H}_{{\rm In}({\ca V})}$, and let $|\Gamma\rangle\in{\ca H}'\otimes{\ca H}_{{\rm In}({\ca V})}$ be the maximally entangled state. By definition, we have
\alg{
&\proj{\Gamma}^{S'S}=({\rm id}^{S'}\otimes({\ca V}^*\circ{\ca V})^S)(\proj{\Gamma})\nn\\
&={\rm Tr}_{R}[(I^{S'}\otimes W)(({\rm id}^{S'}\otimes{\ca V})(\proj{\Gamma}))(I^{S'}\otimes W^\dagger)],\nn
}
which implies that there exists a state $\sigma_0\in{\ca S}({\ca H}^R)$ satisfying
\alg{
&(I^{S'}\otimes W)(({\rm id}^{S'}\otimes{\ca V})(\proj{\Gamma}))(I^{S'}\otimes W^\dagger)\nn\\
&\quad=\proj{\Gamma}^{S'S}\otimes\sigma_0^R.\nn
}
For any pure state $|\varphi\rangle\in{\ca H}_{{\rm In}({\ca V})}$ there exists a supernormalized state $|{\tilde\varphi}\rangle\in{\ca H}'$ such that
\alg{
(\bra{\tilde{\varphi}}^{S'}\otimes I^S)\ket{\Gamma}=|\varphi\rangle^S.\nn
} 
Hence we have
\alg{
&W({\ca V}(\proj{\varphi}))W^\dagger\nn\\
&=(\bra{\tilde{\varphi}}^{S'}\otimes W)(({\rm id}^{S'}\otimes{\ca V})(\proj{\Gamma}))(\ket{\tilde{\varphi}}^{S'}\otimes W^\dagger)\nn\\
&=(\bra{\tilde{\varphi}}^{S'}\otimes I^{SR})(\proj{\Gamma}^{S'S}\otimes\sigma_0^R)(\ket{\tilde{\varphi}}^{S'}\otimes I^{SR})\nn\\
&=\proj{\varphi}^{S}\otimes\sigma_0^R.\nn
}
Noting that any state on $S$ is represented as a probabilistic mixture of pure states, this completes the proof of \req{maria}.\QED
\eprf

\hfill

\section{Proof of Monotones and Bound Sets}\label{app:monotII}

In this Appendix, we prove that the functions introduced in Section \ref{sec:NMmonot} is monotonically nonincreasing under $\Omega$, and that the subsets of states defined in \rSec{boundset} are bound sets under $\Omega$, or equivalently, under any class of operations that comprises $\Omega$. Note that, due to the symmetry of those functions and subsets in exchanging systems $A$ and $B$, monotonicity of the functions and closedness of the subsets under ${\mathbb L}_B$,  ${\mathbb P}_B$ and ${\mathbb Q}_{BE}$ immediately follows from those for ${\mathbb L}_A$,  ${\mathbb P}_A$ and ${\mathbb Q}_{AE}$, respectively.

\subsection{Monotonicity of $I_\downarrow$ and Boundedness of ${\ca S}_I$}

Take arbitrary states $\rho\in{\ca S}_{\rm all}$, $\sigma\in{\ca S}_I$ and operations ${\ca T},{\ca T}_0\in{\mathbb L}_E$ such that ${\ca T}_0(\sigma)\in{\ca S}_I$.

\subsubsection{${\mathbb L}_A$ and ${\mathbb L}_B$}
For any ${\ca E}\in{\mathbb L}_A$, we have
\alg{
I(A:B|E)_{{\ca T}(\rho)}
&\geq (A:B|E)_{({\ca E}\otimes{\ca T})(\rho)}\nn\\
&=I(A:B|E)_{{\ca T}({\ca E}(\rho))},\nn
}
due to the data processing inequality for the conditional quantum mutual information. Taking the infimum over $\ca T$, this implies the monotonicity of $I_\downarrow$. It follows that there exists ${\ca T}_0\in{\mathbb L}_E$ satisfying
\alg{
0=I(A:B|E)_{{\ca T}_0(\sigma)}=I(A:B|E)_{{\ca T}_0({\ca E}(\sigma))},\nn
}
which implies ${\ca E}(\sigma)\in{\ca S}_I$. 

\subsubsection{${\mathbb R}_E$}

By definition, for any ${\ca V}\in{\mathbb R}_E$, there exists an operation ${\ca V}^*\in{\mathbb L}_E$ such that ${\ca V}^*\circ{\ca V}$ is the identity operation on ${\ca S}({\ca H}_{{\rm In}({\ca V})})$. Hence we have
\alg{
I(A:B|E)_{{\ca T}(\rho)}&=I(A:B|E)_{({\ca T}\circ{\ca V}^*)({\ca V}(\rho))}\nn\\
&\geq\inf_{{\ca T}'}I(A:B|E)_{{\ca T}'({\ca V}(\rho))},\nn
}
which implies the monotonicity of $I_\downarrow$ by taking the infimum over all $\ca T$. Denoting ${\ca T}_0\circ{\ca V}^*$ by ${\tilde{\ca T}}_0$, we have 
\alg{
I(A:B|E)_{{\ca T}_0(\sigma)}&=I(A:B|E)_{{\tilde{\ca T}}_0({\ca V}(\sigma))},\nn
}
which implies ${\ca V}(\sigma)\in{\ca S}_I$.

\subsubsection{${\mathbb P}_A$ and ${\mathbb P}_B$}

Recall that the states before and after public communication by Alice are represented by density operators $\rho_i$ and $\rho_f$, respectively, in \req{rhohati} and \req{rhohatf}. We have
\alg{
&I(C_AA:B|E)_{{\ca T}(\rho_i)}\nn\\
&=I(C_A:B|E)_{{\ca T}(\rho_i)}+I(A:B|EC_A)_{{\ca T}(\rho_i)}\nn\\
&\geq I(A:B|EC_A)_{{\ca T}(\rho_i)}\nn\\
&=\sum_mp_mI(A:B|E)_{{\ca T}(\rho_m)}\nn\\
&=I(A:C_BB|C_EE)_{{\ca T}(\rho_f)}.\laeq{pasapasa}
}
It follows that
\alg{
\inf_{\ca T}I(C_AA:B|E)_{{\ca T}(\rho_i)}\geq \inf_{{\ca T}'}I(A:C_BB|E')_{{\ca T}'(\rho_f)}\nn
}
where the infimum in the R.H.S. is taken over all operations from $C_EE$ to $E'$. This implies the monotonicity of $I_\downarrow$. Suppose that $\sigma\in{\ca S}_I$ is decomposed in the form of 
\alg{
\sigma=\sum_mp_m\proj{m}^{C_{\!A}}\otimes\sigma_m^{ABE},\laeq{decsigma}
} 
which is, after the communication by Alice, transformed to
\alg{
\sigma'=\sum_mp_m\proj{m}^{C_{\!B}}\otimes\proj{m}^{C_{\!E}}\otimes\sigma_m^{ABE}.\laeq{decsigmap}
} 
Applying \req{pasapasa} yields
\alg{
0=I(C_AA:B|E)_{{\ca T}_0(\sigma)}\geq I(A:C_BB|C_EE)_{{\ca T}_0(\sigma')},\nn
}
which leads to $\sigma'\in{\ca S}_I$.

\subsubsection{${\mathbb Q}_{AE}$ and ${\mathbb Q}_{BE}$} 
Let $Q$ be a quantum system that is transmitted from Alice to Eve. In the same way as \req{onakaitaii}, for any quantum state $\rho$ on $ABEQ$, we have
\alg{
&I(QA:B|E)_{{\ca T}(\rho)}\nn\\
&=I(Q:B|E)_{{\ca T}(\rho)}+I(A:B|EQ)_{{\ca T}(\rho)}\nn\\
&\geq I(A:B|EQ)_{{\ca T}(\rho)}.\label{eq:onakaitai}
}
It follows that
\alg{
\inf_{\ca T}I(QA:B|E)_{{\ca T}(\rho)}\geq \inf_{{\ca T}'}I(A:B|E')_{{\ca T}'(\rho)}\nn
}
where the infimum in the R.H.S. is taken over all operations from $EQ$ to $E'$. This implies the monotonicity of $I_\downarrow$. From \req{onakaitai}, we have
\alg{
0=I(QA:B|E)_{{\ca T}_0(\sigma)}\geq I(A:B|EQ)_{{\ca T}_0(\sigma)},\nn
}
which implies that ${\ca S}_I$ is closed under ${\mathbb Q}_{BE}$.

\subsubsection{Closedness under Tensor Product}

Consider arbitrary states $\sigma_1,\sigma_2\in{\ca S}_I$ and operations ${\ca T}_1,{\ca T}_2\in{\mathbb L}_E$ such that ${\ca T}_1(\sigma_1),{\ca T}_2(\sigma_2)\in{\ca S}_{\rm Markov}$. Due to the closedness of ${\ca S}_{\rm Markov}$ under tensor product (\rLmm{tensorMarkov}), we have
\alg{
({\ca T}_1\otimes{\ca T}_2)(\rho\otimes\sigma)={\ca T}_1(\rho)\otimes{\ca T}_2(\sigma)\in{\ca S}_{\rm Markov},\nn
}
which implies that ${\ca S}_I$ is closed under tensor product.\hfill$\blacksquare$

\subsection{Monotonicity of $I_\downarrow^*$ and Boundedness of ${\ca S}_I^*$}

Take arbitrary states $\rho\in{\ca S}_{\rm all}$ and $\sigma\in{\ca S}_I^*$, and consider their extensions $\varrho^{ABEF}$ and $\varsigma^{ABEF}$, respectively, such that $\varsigma^{ABF}\in{\ca S}_{\rm Markov}^{A:B|F}$.

\subsubsection{${\mathbb L}_A$ and ${\mathbb L}_B$}

For any ${\ca E}\in{\mathbb L}_A$, an extension of ${\ca E}(\rho)$ is given by ${\ca E}(\vro)$. Hence, due to the data processing inequality for CQMI, we have
\alg{
I(A:B|F)_{\vro}
&\geq I(A:B|F)_{{\ca E}(\vro)}\nn\\
&\geq\inf_{\vro'} I(A:B|F)_{\vro'},\nn
}
where the infimum is taken over all extensions of ${\ca E}(\rho)$. Taking the infimum over all $\vro$, this implies the monotonicity of $I_\downarrow$. It follows that 
\alg{
0=I(A:B|F)_{\varsigma}\geq I(A:B|F)_{{\ca E}(\varsigma)},\nn
}
which implies ${\ca E}(\sigma)\in{\ca S}_I^*$. 

\subsubsection{${\mathbb R}_E$}

For any ${\ca V}\in{\mathbb R}_E$, an extension of ${\ca V}(\rho)$ is given by ${\ca V}(\vro)$. Noting that ${\rm Tr}_{E}[{\ca V}(\varrho^{ABEF})]={\rm Tr}_{E}[\varrho^{ABEF}]$, we have
\alg{
I(A:B|F)_{\vro}
&= I(A:B|F)_{{\ca V}(\vro)}\nn\\
&\geq\inf_{\vro'} I(A:B|F)_{\vro'},\nn
}
where the infimum is taken over all extensions of ${\ca V}(\rho)$. Taking the infimum over all $\vro$, this implies the monotonicity of $I_\downarrow$. It follows that 
\alg{
0=I(A:B|F)_{\varsigma}= I(A:B|F)_{{\ca V}(\varsigma)},\nn
}
which implies ${\ca V}(\sigma)\in{\ca S}_I^*$.

\subsubsection{${\mathbb P}_A$ and ${\mathbb P}_B$}

Recall that the states before and after broadcasting of classical message by Alice are represented by density operators $\rho_i$ in \req{rhohati} and  $\rho_f$ in \req{rhohatf}, respectively. Let $\varrho_i$ be an arbitrary extension of $\rho_i$, and define an extension $\varrho_f$ of $\rho_f$ by \req{rhohatff}.
Denoting by $\ca D$ the dephasing operation on $C_A$ with respect to the basis $\{|m\rangle\}_m$, we have
\alg{
&I(C_AA:B|F)_{\varrho_i}\nn\\
&\geq I(C_AA:B|F)_{{\ca D}(\varrho_i)}\nn\\
&=I(C_A:B|F)_{{\ca D}(\varrho_i)}+I(A:B|FC_A)_{{\ca D}(\varrho_i)}\nn\\
&\geq I(A:B|FC_A)_{{\ca D}(\varrho_i)}\nn\\
&=\sum_mp_mI(A:B|F)_{{\ca D}(\varrho_i)}\nn\\
&=I(A:C_BB|C_{F}F)_{\varrho_f}.\laeq{tontonton}
}
It follows that
\alg{
\inf_{\vro_i}I(C_AA:B|F)_{\varrho_i}=\inf_{\vro'}I(A:C_BB|F)_{\varrho'},\nn
}
where the infimum in the L.H.S is taken over all extensions $\varrho_i^{AC_ABEF}$ of $\rho_i$, while one in the R.H.S. is over all extensions $\varrho_f^{ABC_BEC_EF'}$ of $\rho_f$. This implies the monotonicity of $I_\downarrow^*$. Suppose that $\sigma\in{\ca S}_I^*$ is decomposed in the form of \req{decsigma}, which is transformed to $\sigma'$ given by \req{decsigmap} after the communication by Alice. Define an extension $\varsigma'$ of $\sigma'$ by 
\alg{
{\varsigma}'&:=\sum_mp_m\proj{m}^{C_{\!B}}\otimes\proj{m}^{C_{\!E}}\nn\\
&\quad\quad\quad\quad\quad\quad\otimes\proj{m}^{C_{\!F}}\otimes{\varsigma}_m^{ABEF},\laeq{shitteze}
} 
where
\alg{
{\varsigma}_m^{ABEF}:=p_m^{-1}\bra{m}^{C_A}{\varsigma}^{C_AABEF}\ket{m}^{C_A} \laeq{morii}
}
for each $m$. It follows from \req{tontonton} that
\alg{
0=I(C_AA:B|F)_{\varsigma}\geq I(A:C_BB|C_{F}F)_{\varsigma'},\nn
}
which implies that ${\ca S}_I^*$ is closed under ${\mathbb P}_A$.

\subsubsection{Monotonicity under ${\mathbb Q}_{AE}$ and ${\mathbb Q}_{BE}$} 

Let $Q$ be a quantum system that is transmitted from Alice to Bob. For any quantum state $\rho$ on $ABEQ$ and its extension $\varrho$ on $ABEFQ$, we have
\alg{
I(QA:B|F)_{\vro}\geq I(A:B|F)_{\vro}.\laeq{itakara}
}
It follows that
\alg{
\inf_{\vro}I(QA:B|F)_{\vro}\geq \inf_{\vro}I(A:B|F)_{\vro},\nn
}
where the infimum in both sides is taken over all extensions $\vro$ of $\rho$. This implies the monotonicity of $I_\downarrow^*$. From \req{itakara}, we have
\alg{
0=I(QA:B|F)_{\varsigma}\geq I(A:B|F)_{\varsigma},\nn
}
which implies that ${\ca S}_I^*$ is closed under ${\mathbb Q}_{AE}$.

\subsubsection{Closedness under Tensor Product}

Consider arbitrary states $\sigma_1,\sigma_2\in{\ca S}_I^*$ and extensions $\varsigma_1,\varsigma_2$ thereof, respectively, on $ABEF$ such that $\varsigma_1^{ABF}, \varsigma_2^{ABF}\in{\ca S}_{\rm Markov}^{A:B|F}$. Due to the closedness of ${\ca S}_{\rm Markov}$ under tensor product (\rLmm{tensorMarkov}), we have
\alg{
\varrho^{A_1B_1F_1}\otimes\varsigma^{A_2B_2F_2}\in{\ca S}_{\rm Markov}^{A:B|F}\nn
}
for $A=A_1A_2$, $B=B_1B_2$ and $F=F_1F_2$. Hence ${\ca S}_I^*$ is closed under tensor product.\hfill$\blacksquare$

\subsection{Monotonicity of $I_{sq}$ and Boundedness of ${\ca S}_{\rm sep}$}

Note that any operation in $\Omega$ is regarded as a LOCC (local operations and classical communication) between Alice and Bob if we ignore Eve. The monotonicity of $I_{sq}$ under $\Omega$ immediately follows from the monotonicity of the squashed entanglement $E_{sq}$ under LOCC\cite{christandl04}. In the same way, the closedness of ${\ca S}_{\rm sep}$ under $\Omega$ follows from the fact that any separable state is mapped to another separable state by LOCC.

 Since $\rho^{A_1B_1}\otimes\sigma^{A_2B_2}$ is a separable state between $A_1A_2$ and $B_1B_2$ if both $\rho$ and $\sigma$ are separable, ${\ca S}_{\rm sep}$ is closed under tensor product. \QED

\subsection{Monotonicity of $J_{\downarrow}$ and Boundedness of ${\ca S}_J$}

Take arbitrary states $\rho\in{\ca S}_{\rm all}$ and $\sigma\in{\ca S}_J$, and consider purifications $|\phi_\rho\rangle^{ABEF_AF_B}$ and $|\phi_\sigma\rangle^{ABEF_AF_B}$ thereof, respectively, such that $\phi_\sigma^{ABF_AF_B}\in{\ca S}_{\rm sep}^{AF_A:BF_B}$. Take arbitrary operations ${\ca T},{\ca T}_0\in{\mathbb L}_E$ such that ${\ca T}_0(\phi_\sigma)\in{\ca S}_{\rm Markov}^{AF_A:BF_B|E}$, the existence of which follows from \req{koikoikoi}.

\subsubsection{${\mathbb L}_A$ and ${\mathbb L}_B$} 
For any ${\ca E}\in{\mathbb L}_A$, there exists a quantum system $A_0$ and a linear isometry $\ca U$ from $A$ to $AA_0$ such that a Stinespring dilation of $\ca E$ is given by ${\ca E}(\cdot)={\rm Tr}_{A_0}[{\ca U}(\cdot)]$. Denoting the composite system $A_0F_A$ by ${\tilde F}_A$, the pure state ${\ca U}(\phi_\rho)$ on $AB{\tilde F}_AF_BE$ satisfies ${\rm Tr}_{{\tilde F}_AF_B}[|\phi_\rho\rangle\!\langle\phi_\rho|]={\ca E}(\rho)^{ABE}$. In addition, for any operation $\ca T$ on $E$. we have
\alg{
I(AF_A:BF_B|E)_{{\ca T}(\phi_\rho)}&=I(A{\tilde F}_A:BF_B|E)_{({\ca U}\otimes{\ca T})(\phi_\rho)}\nn\\
&=I(A{\tilde F}_A:BF_B|E)_{{\ca T}({\ca U}(\phi_\rho))}.\laeq{meguriyuku}
}
Hence we have
\alg{
&\inf_{\phi_\rho}\inf_{{\ca T}}I(AF_A:BF_B|E)_{{\ca T}(\phi_\rho)}\nn\\
&\geq \inf_{\phi_{{\ca E}(\rho)}}\inf_{{\ca T}}I(A{\tilde F}_A:BF_B|E)_{{\ca T}(\phi_{{\ca E}(\rho)})}.\nn
}
where the second infimum in the R.H.S. is taken over all purifications $\phi_{{\ca E}(\rho)}$ of ${\ca E}(\rho)$. This implies the monotonicity of $J_{\downarrow}$ under ${\mathbb L}_A$. The monotonicity under ${\mathbb L}_B$ follows along the same line. It follows from \req{meguriyuku} that
\alg{
0=I(AF_A:BF_B|E)_{{\ca T}_0(\phi_\sigma)}=I(A{\tilde F}_A:BF_B|E)_{{\ca T}_0({\ca U}(\phi_\sigma))},\nn
}
which implies ${\ca E}(\sigma)\in{\ca S}_J$.

\subsubsection{${\mathbb R}_E$} 

Let $\ca V$ be an arbitrary reversible operation on $E$, let $E_1$ be an ancillary system and ${\ca U}_1$ be a linear isometry from $E$ to $EE_1$ such that the Stinespring dilation of $\ca V$ is given by ${\ca V}={\rm Tr}_{E_1}\circ{\ca U}_1$. Let ${\ca V}^*$ be an operation on $E$ satisfying ${\ca V}^*\circ{\ca V}={\rm id}$, let $E_2$ be an ancillary system and ${\ca U}_2$ be a linear isometry from $E$ to $EE_2$ such that the Stinespring dilation of ${\ca V}^*$ is given by ${\ca V}^*={\rm Tr}_{E_2}\circ{\ca U}_2$. Define a pure state $|\phi_{\rho}'\rangle$ on $ABEF_AF_BE_1E_2$ by
\alg{
\proj{\phi_{\rho}'}:=({\ca U}_2\circ{\ca U}_1)(\proj{\phi_\rho}).\nn
}
Due to \rLmm{SSrev}, we have
\alg{
{\rm Tr}_{E_1}[|\phi_\rho'\rangle\!\langle\phi_\rho'|]=({\ca U}_2\circ{\ca V})(|\phi_\rho\rangle\!\langle\phi_\rho|)=|\phi_\rho\rangle\!\langle\phi_\rho|\otimes\sigma_0^{E_2}.\nn
}
Denoting by $|\psi_0\rangle$ a purification of $\sigma_0$, it follows that
\alg{
|\phi_{\rho}'\rangle=|\phi_\rho\rangle^{ABEF_AF_B}|\psi_0\rangle^{E_1E_2}.\nn
}
Hence a purification of ${\ca V}(\rho)$ is given by
\alg{
\phi_{{\ca V}(\rho)}^{*ABEF_A{\tilde F}_B}:={\ca U}_2^{-1}(\proj{\phi_\rho}^{ABEF_AF_B}\otm\proj{\psi_0}^{E_1E_2}),\nn
}
where we have denoted the composite system $F_BE_1$ by ${\tilde F}_B$. Define ${\tilde{\ca T}}:=({\ca T}^E\otimes{\rm id}^{E_2})\circ{\ca U}_2$. We have
\alg{
&I(AF_A:BF_B|E)_{{\ca T}(\phi_\rho)}\nn\\
&=I(AF_A:BF_BE_1|EE_2)_{{\ca T}(\phi_\rho)\otimes\psi_0}\nn\\
&=I(AF_A:B{\tilde F}_B|EE_2)_{{\tilde{\ca T}}(\phi_{{\ca V}(\rho)}^*)}.\laeq{lalai}
} 
Thus we obtain
\alg{
&\inf_{\phi_\rho}\inf_{{\ca T}}I(AF_A:BF_B|E)_{{\ca T}(\phi_\rho)}\nn\\
&\geq \inf_{\phi_{{\ca V}(\rho)}}\inf_{{\ca T}'}I(AF_A:BF_B|E')_{{\ca T}'(\phi_{{\ca V}(\rho)})},\nn
}
where the first infimum in the R.H.S. is taken with respect to all operations ${\ca T}'$ from $E$ to $E'$, and the second one is over all purifications $\phi_{{\ca V}(\rho)}$ of ${\ca V}(\rho)$. This implies the monotonicity under ${\mathbb R}_E$. It follows from \req{lalai} that
\alg{
I(AF_A:BF_B|E)_{{\ca T}_0(\phi_\sigma)}=I(AF_A:B{\tilde F}_B|EE_2)_{{\tilde{\ca T}}_0(\phi_{{\ca V}(\sigma)}^*)},\nn
} 
which yields ${\ca V}(\sigma)\in{\ca S}_J$.

\subsubsection{${\mathbb P}_A$ and ${\mathbb P}_B$}

Recall that the states before and after public communication by Alice are represented by density operators $\rho_i$ and $\rho_f$ in \req{rhohati} and \req{rhohatf}, and that purifications of those states are given by $\phi_{\rho_i}^*$ in \req{watashinara} and $\phi_{\rho_f}^*$ in \req{anatanara}, respectively. Let ${\ca D^A}$ and ${\ca D^E}$ be the dephasing operation on $C_A$ and $C_E$ with respect to the basis $\{|m\rangle\}$. Due to Uhlmann's theorem\cite{uhlmann1976transition}, for any purification $\phi_{\rho_i}$ of $\rho_i$, there exists a linear isometry $\ca W$ from $C_{F}F$ to $F_AF_B$ such that ${\ca W}(\proj{\phi_{\rho_i}^*})=\proj{\phi_{\rho_i}}$. We have
\alg{
&I(C_AAF_A:BF_B|E)_{{\ca T}(\phi_{\rho_i})}\nn\\
&=I(C_AAF_A:BF_B|E)_{({\ca W}\otimes{\ca T})(\phi_{\rho_i}^*)}\nn\\
&\geq I(C_AAF_A:BF_B|E)_{({\ca D}^A\otimes{\ca W}\otimes{\ca T})(\phi_{\rho_i}^*)}\nn\\
&=I(C_A:BF_B|E)_{({\ca D}^A\otimes{\ca W}\otimes{\ca T})(\phi_{\rho_i}^*)}\nn\\
&\quad\quad+I(AF_A:BF_B|EC_A)_{({\ca D}^A\otimes{\ca W}\otimes{\ca T})(\phi_{\rho_i}^*)}\nn\\
&\geq I(AF_A:BF_B|EC_A)_{({\ca D}^A\otimes{\ca W}\otimes{\ca T})(\phi_{\rho_i}^*)}\nn\\
&=\sum_mp_mI(AF_A:BF_B|E)_{({\ca W}\otimes{\ca T})(\phi_{\rho_m})}\nn\\
&=I(AF_A:C_BBF_B|M_{E}E)_{({\ca D}^A\otimes{\ca W}\otimes{\ca T})(\phi_{\rho_f}^*)}\nn\\
&=I(AF_A:C_BBF_B|M_{E}E)_{({\ca W}\otimes{\ca D}^E\otimes{\ca T})(\phi_{\rho_f}^*)}\nn\\
&= I(AF_A:C_BBF_B|M_{E}E)_{{\tilde{\ca T}}({\ca W}(\phi_{\rho_f}^*))},\label{eq:cloudd}
}
where we defined ${\tilde T}:={\ca D}^E\otimes{\ca T}$ in the last line. Noting that ${\ca W}(\phi_{\rho_f})$ is a purification of $\rho_f$, it follows that
\alg{
&\inf_{\phi_{\rho_i}}\inf_{{\ca T}}I(AF_A:BF_B|E)_{{\ca T}(\phi_{\rho_i})}\nn\\
&\geq \inf_{\phi_{\rho_f}}\inf_{{\ca T}'}I(AF_A:C_BBF_B|E')_{{\ca T}'(\phi_{\rho_f})},\nn
}
where the infimum in the R.H.S. is taken over all operations ${\ca T}'$ from $M_{E}E$ to $E'$ and all purifications $\phi_{\rho_f}$ of $\rho_f$. This implies the monotonicity of $J_\downarrow^*$. Suppose that $\sigma\in{\ca S}_J$ is decomposed in the form of \req{decsigma}, which is transformed to $\sigma'$ in \req{decsigmap} after communication by Alice. Denoting by $|\phi_{\sigma_m}\rangle$ a purification of $\sigma_m$ for each $m$, we obtain the following purifications of $\sigma$ and $\sigma'$, respectively:
\alg{
&|\phi_{\sigma}^*\rangle:=\sum_m\sqrt{p_m}|m\rangle^{C_A}|m\rangle^{C_{F}}|\phi_{\sigma_m}\rangle^{ABEF},\nn\\
&|\phi_{\sigma'}^*\rangle:=\sum_m\sqrt{p_m}|m\rangle^{C_B}|m\rangle^{C_E}|m\rangle^{C_{F}}|\phi_{\sigma_m}\rangle^{ABEF}.\nn
}
From \req{cloudd}, we obtain 
\alg{
&0\geq I(C_AAF_A:BF_B|E)_{({\ca W}\otimes{\ca T}_0)(\phi_{\sigma}^*)}\nn\\
&\geq I(AF_A:C_BBF_B|M_{E}E)_{{\tilde{\ca T}}({\ca W}(\phi_{\sigma'}^*))}.\nn
}
Noting that ${\ca W}(\phi_{\sigma'}^*)$ is a purification of $\sigma'$, the above inequality implies $\sigma'\in{\ca S}_J$.

\subsubsection{${\mathbb Q}_{AE}$ and ${\mathbb Q}_{BE}$} 

Let $Q$ be a quantum system that is transmitted from Alice to Bob. In the same way as (\ref{eq:onakaitai}), for any quantum state $\rho$ on $ABEQ$ and for any operation $\ca T$ on $E$, we have
\alg{
&I(QAF_A:BF_B|E)_{{\ca T}(\phi_\rho)}\nn\\
&=I(Q:BF_B|E)_{{\ca T}(\phi_\rho)}+I(AF_A:BF_B|EQ)_{{\ca T}(\phi_\rho)}\nn\\
&\geq I(AF_A:BF_B|EQ)_{{\ca T}(\phi_\rho)}.\laeq{miramira}
}
It follows that
\alg{
&\inf_{\phi_\rho}\inf_{\ca T}I(QAF_A:BF_B|E)_{{\ca T}(\phi_\rho)}\nn\\
&\geq \inf_{\phi_\rho}\inf_{{\ca T}'}I(AF_A:BF_B|E')_{{\ca T}'(\phi_\rho)}\nn
}
where the second infimum in the R.H.S. is taken over all operations from $EQ$ to $E'$. The monotonicity under ${\mathbb Q}_{BE}$ follows along the same line. From \req{miramira}, we have
\alg{
0=I(QAF_A:BF_B|E)_{{\ca T}_0(\phi_\sigma)}\geq I(AF_A:BF_B|EQ)_{{\ca T}_0(\phi_\sigma)},\nn
}
which implies that ${\ca S}_J$ is closed under ${\mathbb Q}_{AE}$.

\subsubsection{Closedness under Tensor Product}

Consider arbitrary states $\sigma_1,\sigma_2\in{\ca S}_J$ and purifications $\phi_{\sigma_1}$, $\phi_{\sigma_2}$ thereof, respectively, on $ABEF_AF_B$ such that $\phi_{\sigma_1}^{ABF_AF_B}, \phi_{\sigma_2}^{ABF_AF_B}\in{\ca S}_{\rm sep}^{AF_A:BF_B}$. It is straightforward to verify that we have
\alg{
\phi_{\sigma_1}^{A_1F_{A1}B_1F_{B_1}}\otimes\phi_{\sigma_2}^{A_2F_{A2}B_2F_{B_2}}\in{\ca S}_{\rm sep}^{AF_A:BF_B}\nn
}
for $A=A_1A_2$, $B=B_1B_2$ and $F_A=F_{A1}F_{A2}$, $F_B=F_{B1}F_{B2}$. Hence ${\ca S}_J$ is closed under tensor product. \hfill\QED

\subsection{Monotonicity of $J_{\downarrow}^*$ and Boundedness of ${\ca S}_J^*$}

Take arbitrary states $\rho\in{\ca S}_{\rm all}$ and $\sigma\in{\ca S}_J^*$, let $\varrho^{ABEF}$ and $\varsigma^{ABEF}$ be extensions thereof, respectively, and let ${\ca W}$ and ${\ca W}_0$ be linear isometries from $E$ to $E_AE_B$  such that such that ${\ca W}_0(\varsigma)\in{\ca S}_{\rm sep}^{AE_A:BE_B}$.

\subsubsection{${\mathbb L}_A$ and ${\mathbb L}_B$} 

By the monotonicity of the conditional quantum mutual information, for any ${\ca E}\in{\mathbb L}_A$ we have
\alg{
&I(AE_A:BE_B|F)_{{\ca W}(\varrho)}\nn\\
&\geq I(AE_A:BE_B|F)_{({\ca E}\otimes{\ca W})(\varrho)}\nn\\
&= I(AE_A:BE_B|F)_{{\ca W}({\ca E}(\varrho))}.\laeq{ennkenn}
}
Noting that ${\ca E}(\varrho)$ is an extension of ${\ca E}(\rho)$, it follows that
\alg{
&\inf_{\varrho}\inf_{\ca W}I(AE_A:BE_B|F)_{{\ca W}(\varrho)}\nn\\
&\geq \inf_{\varrho'}\inf_{\ca W}I(AE_A:BE_B|F)_{{\ca W}(\varrho')},\nn
}
where the second infimum in the R.H.S. is taken over all extensions $\varrho'$ of ${\ca E}(\rho)$. Thus we obtain the monotonicity of $J_\downarrow^*$. Inequality \req{ennkenn} implies
\alg{
0=I(AE_A:BE_B|F)_{{\ca W}_0(\varsigma)}\geq I(AE_A:BE_B|F)_{{\ca W}_0({\ca E}(\varsigma))},\nn
}
which yields ${\ca E}(\sigma)\in{\ca S}_J^*$.

\subsubsection{${\mathbb R}_E$}

Suppose ${\ca V}\in{\mathbb R}_E$. Due to \rLmm{SSrev}, there exists a state $\sigma_0$ and a linear isometry $\ca U$ from $E$ to $EE_0$ such that a Stinespring dilation of ${\ca V}^*$ is given by ${\ca V}^*(\cdot)={\rm Tr}_{E_0}[{\ca U}(\cdot)]$, and that 
\alg{
{\ca U}({\ca V}(\tau))=\tau\otimes\sigma_0\label{eq:samishii}
}
holds for any $\tau$. Denote $E_BE_0$ by ${\tilde E}_B$, and define a linear isometry $\tilde{\ca W}$ from $E$ to $E_A{\tilde E}_B$ by $\tilde{\ca W}:=({\ca W}\otimes{\rm id}^{E_0})\circ{\ca U}$. Due to (\ref{eq:samishii}), we have
\alg{
I(AE_A:BE_B|F)_{{\ca W}(\varrho)}=I(AE_A:B{\tilde E}_B|F)_{{\tilde{\ca W}}({\ca V}(\varrho))}.\laeq{nanare}
}
Noting that ${\ca V}(\varrho)$ is an extension of ${\ca V}(\rho)$, we have
\alg{
&\inf_{\varrho}\inf_{\ca W}I(AE_A:BE_B|F)_{{\ca W}(\varrho)}\nn\\
&\geq \inf_{\varrho'}\inf_{{\ca W}'}I(AE_A:BE_B|F)_{{{\ca W}'}(\varrho')},\nn
}
where the infimum in the R.H.S. is taken with respect to all extensions $\varrho'$ of $\rho$ and linear isometries ${\ca W}'$ of $E$ to $E_AE_B$. This implies the monotonicity of $J_{\downarrow}^*$. From \req{nanare}, we obtain 
\alg{
0=I(AE_A:BE_B|F)_{{\ca W}_0(\varsigma)}=I(AE_A:B{\tilde E}_B|F)_{{\tilde{\ca W}_0}({\ca V}(\varsigma))},\nn
}
which leads to ${\ca V}(\sigma)\in{\ca S}_J^*$.

\subsubsection{${\mathbb P}_A$ and ${\mathbb P}_B$}

Recall that the states before and after public communication by Alice are represented by density operators $\rho_i$ and $\rho_f$ in \req{rhohati} and \req{rhohatf}, respectively. Let $\varrho_i$ be an arbitrary extension of $\rho_i$, and define an extension $\varrho_f$ of $\rho_f$ by \req{rhohatff}.
Denoting by $\ca D$ the dephasing operation on $C_A$ with respect to the basis $\{|m\rangle\}_m$, we have
\alg{
&I(C_AAE_A:BE_B|F)_{{\ca W}({\varrho_i})}\nn\\
&\geq I(C_AAE_A:BE_B|F)_{({\ca D}\otimes{\ca W})(\varrho_i)}\nn\\
&=I(C_A:BE_B|F)_{({\ca D}\otimes{\ca W})(\varrho_i)}\nn\\
&\quad\quad+I(AE_A:BE_B|FC_A)_{({\ca D}\otimes{\ca W})(\varrho_i)}\nn\\
&\geq I(AE_A:BE_B|FC_A)_{({\ca D}\otimes{\ca W})(\varrho_i)}\nn\\
&=\sum_mp_mI(AE_A:BE_B|F)_{{\ca W}(\varrho_m)}\nn\\
&=I(AE_A:C_BBE_BC_E|C_{F}F)_{{\ca W}({\varrho_f})}\nn\\
&=I(AE_A:C_BB{\tilde E}_B|C_{F}F)_{{\tilde{\ca W}}({\varrho_f})}.\label{eq:cloudy}
}
Here, we have denoted $E_BC_E$ by ${\tilde E}_B$, and denoted a linear isometry ${\ca W}\otm{\rm id}^{C_E}$ from $EC_E$ to $E_A{\tilde E}_B$ by $\tilde{\ca W}$. Hence we obtain
\alg{
&\inf_{\varrho_i}\inf_{\ca W}I(C_AAE_A:BE_B|F)_{{\ca W}({\varrho_i})}\nn\\
&\geq \inf_{\varrho_f'}\inf_{{\ca W}'}I(AE_A:C_BBE_B|F')_{{\ca W}'({\varrho_f}')},\nn
}
where the infimum in the second line is taken over all linear isometries ${\ca W}'$ from $EC_E$ to $E_AE_B$ and all extensions $\varrho_f'$ of $\rho_f$. The monotonicity under ${\mathbb P}_B$ follows along the same line. Suppose that $\sigma\in{\ca S}_J^*$ is decomposed in the form of \req{decsigma}, which is transformed to $\sigma'$ given by \req{decsigmap} after the communication by Alice. Define an extension $\varsigma'$ of $\sigma'$ by \req{shitteze} and \req{morii}. Inequality \req{cloudy} yields
\alg{
&0=I(C_AAE_A:BE_B|F)_{{\ca W}_0({\varsigma})}\nn\\
&\geq I(AE_A:C_BB{\tilde E}_B|C_{F}F)_{{\tilde{\ca W}}_0(\varsigma')},\nn
}
which leads to $\sigma'\in{\ca S}_J^*$.

\subsubsection{${\mathbb Q}_{AE}$ and ${\mathbb Q}_{BE}$} 
Let $Q$ be a quantum system that is transmitted from Alice to Bob, and define an operation $\tilde{\ca W}$ from $EQ$ to $E_AE_BQ$ by $\tilde{\ca W}:={\ca W}\otimes{\rm id}^Q$. Denoting the composite system $E_AQ$ by ${\tilde E}_A$, we have
\alg{
\!\!\!\!I((AQ)E_A:BE_B|F)_{{\ca W}(\varrho)}=I(A{\tilde E}_A:BE_B|F)_{{\tilde{\ca W}}(\varrho)}.\laeq{itakarara}
}
Hence we obtain
\alg{
&\inf_{\varrho}\inf_{{\ca W}}I((AQ)E_A:BE_B|F)_{{\ca W}(\varrho)}\nn\\
&\geq\inf_{\varrho'}\inf_{{\ca W}'}I(A{\tilde E}_A:BE_B|F)_{{\ca W}'(\varrho')},\nn
}
where the second infimum in the R.H.S. is taken over all linear isometries from $EQ$ to ${\tilde E}_AE_B$. This implies the monotonicity of $J_\downarrow^*$. From \req{itakarara}, we have
\alg{
0=I((AQ)E_A:BE_B|F)_{{\ca W}_0(\varsigma)}=I(A{\tilde E}_A:BE_B|F)_{{\tilde{\ca W}}_0(\varsigma)},\nn
}
which implies that ${\ca S}_J^*$ is closed under ${\mathbb Q}_{AE}$.

\subsubsection{Closedness under Tensor Product}

Consider arbitrary states $\sigma_1,\sigma_2\in{\ca S}_J^*$ and linear isometries ${\ca W}_1$, ${\ca W}_2$ of $E$ to $E_AE_B$ such that ${\ca W}_1(\sigma_1), {\ca W}_2(\sigma_2)\in{\ca S}_{\rm sep}^{AE_A:BE_B}$. It is straightforward to verify that
\alg{
\varrho^{A_1B_1}\otimes\varsigma^{A_2B_2}\in{\ca S}_{\rm sep}^{A_1A_2:B_1B_2}.\nn
}
Hence ${\ca S}_J^*$ is closed under tensor product. \hfill\QED

\subsection{Monotonicity of $D_{\rm rec}^B$ under $\Omega^\rightarrow$}\label{app:monotrer}

\subsubsection{Monotonicity under ${\mathbb L}_A$ and ${\mathbb L}_B$} 
Suppose ${\ca E}\in{\mathbb L}_A$. Due to the monotonicity of the quantum relative entropy, we have
\alg{
&\inf_{\ca R}D(\rho^{ABE}\|{\ca R}(\rho^{AE}))\nn\\
&\geq\inf_{{\ca R}}D({\ca E}(\rho^{ABE})\|({\ca E}\otimes{\ca R})(\rho^{AE}))\nn\\
&=\inf_{{\ca R}}D({\ca E}(\rho^{ABE})\|{\ca R}({\ca E}(\rho^{AE}))).\nn
}
where the infimum is taken over all quantum operations ${\ca R}$ from $E$ to $BE$.  This implies the monotonicity under local operations by Alice.  To prove the monotonicity under Bob's operation, suppose ${\ca E}\in{\mathbb L}_B$. We have
\alg{
&\inf_{\ca R}D(\rho^{ABE}\|{\ca R}(\rho^{AE}))\nn\\
&\geq\inf_{{\ca R}}D({\ca E}(\rho^{ABE})\|({\ca E}\circ{\ca R})(\rho^{AE}))\nn\\
&\geq\inf_{{\ca R}'}D({\ca E}(\rho^{ABE})\|{\ca R}'(\rho^{AE})),\nn
} 
where infimum in the last line is taken with respect to all quantum operations ${\ca R}'$ from $E$ to $BE$. Noting that we have
\alg{
\rho^{AE}={\rm Tr}_B[\rho^{ABE}]={\rm Tr}_B[{\ca E}(\rho^{ABE})],\nn
}
this implies the monotonicity under ${\mathbb L}_B$.

\subsubsection{Monotonicity under ${\mathbb R}_E$} 
Suppose ${\ca V}\in{\mathbb R}_E$. Then
\alg{
&\inf_{\ca R}D(\rho^{ABE}\|{\ca R}(\rho^{AE}))\nn\\
&\geq\inf_{\ca R}D({\ca V}(\rho^{ABE})\|({\ca V}\circ{\ca R})(\rho^{AE}))\nn\\
&=\inf_{\ca R}D({\ca V}(\rho^{ABE})\|({\ca V}\circ{\ca R}\circ{\ca V}^{-1})({\ca V}(\rho^{AE}))),\nn\\
&\geq\inf_{{\ca R}'}D({\ca V}(\rho^{ABE})\|{\ca R}'({\ca V}(\rho^{AE}))),\nn
}
where the infimum is taken over all quantum operations ${\ca R}'$ from $E$ to $BE$. This implies $D_{\rm rec}^B(\rho)\geq D_{\rm rec}^B({\ca V}(\rho))$.

\subsubsection{Monotonicity under ${\mathbb P}_A$}
 Consider states $\rho_i$ and $\rho_f$ defined by (\ref{eq:rhohati}) and (\ref{eq:rhohatf}), respectively. Let ${\ca R}$ be an arbitrary quantum operation from $E$ to $BE$. We have
\alg{
{\ca R}(\rho_i^{C_AAE})=\sum_mp_m\proj{m}^{C_A}\otimes{\ca R}(\rho_m^{AE}),\nn
}
which yields
\alg{
D(\rho_i^{C_AABE}\|{\ca R}(\rho_i^{C_AAE}))=\sum_mp_mD(\rho_m^{ABE}\|{\ca R}(\rho_m^{AE})).\label{eq:namacha1}
}
Define a quantum operation ${\tilde{\ca R}}$ from $C_EE$ to $C_BBC_EE$ by
\alg{
{\tilde{\ca R}}(\proj{m}^{C_E}\otimes\tau^{E})=\proj{m}^{C_B}\otimes\proj{m}^{C_E}\otimes{\ca R}(\tau).\nn
}
It is straightforward to verify that
\alg{
{\tilde{\ca R}}(\rho_f^{AC_EE})=\sum_mp_m\proj{m}^{C_B}\otimes\proj{m}^{C_E}\otimes{\ca R}(\rho_m^{AE}),\nn
}
and thus we have
\alg{
D(\rho_f^{AC_BBC_EE}\|{\tilde{\ca R}}(\rho_f^{AC_EE}))=\sum_mp_mD(\rho_m^{ABE}\|{\ca R}(\rho_m^{AE})).\nn
}
Combining this with (\ref{eq:namacha1}), we obtain
\alg{
&D(\rho_i^{C_AABE}\|{\ca R}(\rho_i^{C_AAE}))\nn\\
&=D(\rho_f^{AC_BBC_EE}\|{\tilde{\ca R}}(\rho_f^{AC_EE}))\nn\\
&\geq\inf_{{\ca R}'}D(\rho_f^{AC_BBC_EE}\|{\ca R}'(\rho_f^{AC_EE})),\nn
}
where the infimum in the last line is taken over all operations ${\ca R}'$ from $C_EE$ to $C_BBC_EE$. This implies the monotonicity under ${\mathbb P}_A$.

\subsubsection{Monotonicity under ${\mathbb Q}_{AE}$ and ${\mathbb Q}_{BE}$}
Let $Q$ be a quantum system that is transmitted from Alice to Eve, and let $\rho$ be a quantum state on $ABEQ$. We have
\alg{
&\inf_{{\ca R}}D(\rho^{ABEQ}\|{\ca R}({\rho}^{AEQ})) \nn\\
&\quad\geq \inf_{{\ca R}'}D(\rho^{ABEQ}\|{\ca R}'({\rho}^{AEQ})),\nn
}
where the infimum in the L.H.S. is taken over all quantum operations from $E$ to $BE$, and one in the R.H.S. over all operations from $EQ$ to $BEQ$. This implies the monotonicity under ${\mathbb Q}_{AE}$. Suppose now that $Q$ is a quantum system that is transmitted from Bob to Eve. We have
\alg{
&\inf_{{\ca R}}D(\rho^{ABEQ}\|{\ca R}({\rho}^{AE})) \nn\\
&\quad\geq \inf_{{\ca R}'}D(\rho^{ABEQ}\|{\ca R}'({\rho}^{AEQ})),\nn
}
where the infimum in the L.H.S. is in this case taken over all quantum operations from $E$ to $BEQ$. This implies the monotonicity under ${\mathbb Q}_{BE}$. \hfill$\blacksquare$

\hfill

\section{Proof of (\ref{eq:euston})}\label{app:prforaora}

In this appendix, we prove Relation (\ref{eq:euston}) after presenting basic calculations of Pauli operators. 

\subsection{Calculation of Pauli operators}

We assume that $k,l,m,n,p,q$ take values in $\{0,1\}$. From $\sigma_x\sigma_z=-\sigma_z\sigma_x$, we have
\alg{
\sigma_x^k\sigma_z^l\sigma_x^p\sigma_z^q
&=(-1)^{lp}\sigma_x^k\sigma_x^p\sigma_z^l\sigma_z^q\nn\\
&=(-1)^{lp}\sigma_x^p\sigma_x^k\sigma_z^q\sigma_z^l\nn\\
&=(-1)^{lp\oplus kq}\sigma_x^p\sigma_z^q\sigma_x^k\sigma_z^l,\nn
}
where $\oplus$ denoted summation modulo $2$. Therefore, noting that
\alg{
&(I\otm\sigma_x)|\Phi_{00}\rangle=\frac{1}{\sqrt{2}}\left(\ket{01}+\ket{10}\right)=(\sigma_x\otimes I)|\Phi_{00}\rangle\nn\\
&(I\otm\sigma_z)|\Phi_{00}\rangle=\frac{1}{\sqrt{2}}\left(\ket{00}-\ket{11}\right)=(\sigma_z\otimes I)|\Phi_{00}\rangle,\nn
} 
we have
\alg{
&(\sigma_x^k\sigma_z^l\otimes \sigma_x^m\sigma_z^n)|\Phi_{pq}\rangle\nn\\
&=(\sigma_x^k\sigma_z^l\otimes \sigma_x^m\sigma_z^n)(\sigma_x^p\sigma_z^q\otimes I)|\Phi_{00}\rangle\nn\\
&=(\sigma_x^k\sigma_z^l\sigma_x^p\sigma_z^q\otimes \sigma_x^m\sigma_z^n)|\Phi_{00}\rangle\nn\\
&=(-1)^{lp\oplus kq}(\sigma_x^p\sigma_z^q\sigma_x^k\sigma_z^l\otimes \sigma_x^m\sigma_z^n)|\Phi_{00}\rangle\nn\\
&=(-1)^{lp\oplus kq}(\sigma_x^p\sigma_z^q\sigma_x^k\sigma_z^l\sigma_z^n\sigma_x^m\otimes I)|\Phi_{00}\rangle\nn\\
&=(-1)^{lp\oplus kq\oplus m(l\oplus n)}(\sigma_x^p\sigma_z^q\sigma_x^{k\oplus m}\sigma_z^{l\oplus n}\otimes I)|\Phi_{00}\rangle\laeq{ohanakirei}\\
&=(-1)^{lp\oplus kq\oplus m(l\oplus n)}(\sigma_x^p\sigma_z^q\otimes I)|\Phi_{k\oplus m,l\oplus n}\rangle.\laeq{irrurura}
}
It follows that
\alg{
(\sigma_x^k\sigma_z^l\otimes \sigma_x^k\sigma_z^l)|\Phi_{pq}\rangle&=(-1)^{lp\oplus kq}(\sigma_x^p\sigma_z^q\otimes I)|\Phi_{00}\rangle\nn\\
&=(-1)^{lp\oplus kq}|\Phi_{pq}\rangle.\label{eq:sssusu}
}
Thus we have
\alg{
&(\sigma_z^{l\oplus n}\otimes I)(\sigma_z^{l}\otimes\sigma_z^{n})|\Phi_{pq}\rangle\nn\\
&=(\sigma_z^{n}\otimes\sigma_z^{n})|\Phi_{pq}\rangle\nn\\
&=(-1)^{np}|\Phi_{pq}\rangle,\nn
}
in addition to
\alg{
&(\sigma_x^{k\oplus m}\otimes I)(\sigma_x^{k}\otimes\sigma_x^{m})|\Phi_{pq}\rangle\nn\\
&=(\sigma_x^{m}\otimes\sigma_x^{m})|\Phi_{pq}\rangle\nn\\
&=(-1)^{mq}|\Phi_{pq}\rangle.\nn
}

Define $|{\bf\Phi}\rangle^{ABE_1E_2}:=|\Phi_{00}\rangle^{AE_1}|\Phi_{00}\rangle^{BE_2}$. A Schmidt decomposition of this state between $AB$ and $E_1E_2$ is given by
\alg{
|{\bf\Phi}\rangle^{ABE_1E_2}=\frac{1}{2}\sum_{p,q=0,1}|\Phi_{pq}\rangle^{AB}|\Phi_{pq}\rangle^{E_1E_2}.\nn
}
Define also
\alg{
|{\bf\Phi}_{kl}\rangle^{ABE_1E_2}:=|\Phi_{kl}\rangle^{AE_1}|\Phi_{kl}\rangle^{BE_2}.\laeq{furu}
}
Applying (\ref{eq:sssusu}), we have
\alg{
&|{\bf\Phi}_{kl}\rangle^{ABE_1E_2}\nn\\
&=((\sigma_x^k\sigma_z^l\otimes I)|\Phi_{00}\rangle)^{AE_1}\otimes((\sigma_x^k\sigma_z^l\otimes I)|\Phi_{00}\rangle)^{BE_2}\nn\\
&=((\sigma_x^k\sigma_z^l)^A\otimes(\sigma_x^k\sigma_z^l)^B\otimes I^{E_1E_2})|{\bf\Phi}\rangle^{ABE_1E_2}\nn\\
&=\frac{1}{2}\sum_{p,q=0,1}((\sigma_x^k\sigma_z^l\otimes\sigma_x^k\sigma_z^l)|\Phi_{pq}\rangle)^{AB}|\Phi_{pq}\rangle^{E_1E_2}\nn\\
&=\frac{1}{2}\sum_{p,q=0,1}(-1)^{lp\oplus kq}|\Phi_{pq}\rangle^{AB}|\Phi_{pq}\rangle^{E_1E_2}.\laeq{furuu}
}
Thus we obtain
\alg{
&((\sigma_x^{k}\sigma_z^{l})^A\otimes(\sigma_x^{m}\sigma_z^{n})^B\otimes(\sigma_x^{k\oplus m}\sigma_z^{l\oplus n})^{E_1}\otimes I^{E_2})|{\bf\Phi}\rangle^{ABE_1E_2}\nn\\
&=((\sigma_x^{k}\sigma_z^{l})^A\otimes(\sigma_x^{k\oplus m}\sigma_z^{l\oplus n})^{E_1})|\Phi_{00}\rangle^{AE_1}\nn\\
&\quad\quad\quad\otimes((\sigma_x^{m}\sigma_z^{n})^B\otimes I^{E_2})|\Phi_{00}\rangle^{BE_2}\nn\\
&=(-1)^{n(k\oplus m)}((\sigma_x^{m}\sigma_z^{n})^A\otimes I^{E_1})|\Phi_{00}\rangle^{AE_1}\nn\\
&\quad\quad\quad\otimes((\sigma_x^{m}\sigma_z^{n})^B\otimes I^{E_2})|\Phi_{00}\rangle^{BE_2}\nn\\
&=(-1)^{n(k\oplus m)}|\Phi_{mn}\rangle^{AE_1}|\Phi_{mn}\rangle^{BE_2}\nn\\
&=(-1)^{n(k\oplus m)}|{\bf\Phi}_{mn}\rangle^{ABE_1E_2}\nn\\
&=\frac{1}{2}\sum_{p,q=0,1}(-1)^{np\oplus mq\oplus n(k\oplus m)}|\Phi_{pq}\rangle^{AB}|\Phi_{pq}\rangle^{E_1E_2},\label{eq:shigechii}
}
where we used \req{ohanakirei} in the second line.

\subsection{$\Psi_{\rm I\!I}\sim\Psi_{\rm I\!I}^*$}

Suppose Alice and Bob apply the Hadamard operation $H$ on their qubit, respectively, and perform the dephasing operation with respect to the basis $\{\ket{0},\ket{1}\}$. Noting that we have
\alg{
&(H^A\otimes H^B\otimes I^E)|\psi_p\rangle^{ABE}\nn\\
&\quad=\frac{1}{2}((\sigma_z^p)^A\otimes I^{BE})\left(\ket{000}+\ket{110}+\ket{011}+\ket{101}\right),\nn
}
the state $\psi_p$ defined by \req{watatame} is transformed to $\Psi_{\rm I\!I}^*$ regardless of the value of $p$. Recalling (\ref{eq:schwartz}), this implies $\Psi_{\rm I\!I}\succ\Psi_{\rm I\!I}^*$.

Let $\{A,A'\}$, $\{B,B'\}$ and $\{E_1,E_2,E'\}$ be qubit systems that are possessed by Alice, Bob and Eve, respectively. By quantum communication from Alice and Bob to Eve, the state $|{\bf\Phi}\rangle^{ABE_1E_2}=|\Phi_{00}\rangle^{AE_1}|\Phi_{00}\rangle^{BE_2}$ can be prepared. Consider the following protocol that is to be performed on the state $\Psi_{\rm I\!I}^{*A'B'E'}\otimes{\bf\Phi}^{ABE_1E_2}$:
\begin{itemize}
\item[(P2)] Alice, Bob and Eve locally perform the following controlled-unitary operations on their qubits:
\alg{
&U_A:=\sum_{l=0,1}\proj{l}^{A'}\!\otimes(\sigma_z^{l})^A,\nn\\
&U_B:=\sum_{n=0,1}\proj{n}^{B'}\!\otimes(\sigma_z^{n})^B,\nn\\
&V_E:=\sum_{s=0,1}\proj{s}^{E'}\otimes(\sigma_z^{s}\otimes I)^{E_1E_2}.\nn
}
Then Alice and Bob discard systems $A'$ and $B'$, and Eve performs the basis transformation $\ket{\Phi_{pq}}\rightarrow|p,q\rangle$.
\end{itemize}
Due to (\ref{eq:shigechii}), we have
\alg{
&\langle l,n|^{A'B'}(U_A\otimes U_B\otimes V_E)|{l,n,l\oplus n}\rangle^{A'B'E'}|{\bf\Phi}\rangle^{ABE_1E_2}\nn\\
&=|{l\oplus n}\rangle^{E'}\otimes((\sigma_z^{l})^A\otimes(\sigma_z^{n})^B\otimes(\sigma_z^{l\oplus n})^{E_1}\otimes I^{E_2})|{\bf\Phi}\rangle\nn\\
&=|{l\oplus n}\rangle^{E'}\otimes\frac{1}{2}\sum_{p,q=0,1}(-1)^{np}|\Phi_{pq}\rangle^{AB}|\Phi_{pq}\rangle^{E_1E_2}\label{eq:raashii}
}
for each value of $l$ and $n$. Noting that $\Psi_{\rm I\!I}^*$ is represented as
\alg{
\Psi_{\rm I\!I}^{*A'B'E'}=\frac{1}{4}\sum_{l,n=0,1}\proj{l}^{A'}\otimes\proj{n}^{B'}\otimes\proj{l\oplus n}^{E'},\nn
}
the state after discarding of $A'$ and $B'$ by Alice and Bob is equal to the probabilistic mixture of (\ref{eq:raashii}) over $l,n\in\{0,1\}$ with respect to the uniform distribution, which is given by
\alg{
\!\Psi_{\rm I\!I}'^{ABE_1E_2}:=\frac{1}{4}\sum_{p,q,q'=0,1}|\Phi_{pq}\rangle\!\langle\Phi_{pq'}|^{AB}\otimes|\Phi_{pq}\rangle\!\langle\Phi_{pq'}|^{E_1E_2}.\nn
}
It is straightforward to verify that the above state is transformed to $\Psi_{\rm I\!I}$ by the basis transformation $\ket{\Phi_{pq}}\rightarrow|p,q\rangle$.

\subsection{$\Psi_{\rm I\!I}^{*\otimes 2}\sim\Phi_{\rm I\!I\!I}$}

Consider the following protocol that is to be performed on the state $\Psi_{\rm I\!I}^{*A'B'E'}\otimes\Psi_{\rm I\!I}^{*A''B''E''}\otimes{\bf\Phi}^{ABE_AE_B}$:
\begin{itemize}
\item[(P3)] Alice, Bob and Eve locally perform the following controlled-unitary operations on their qubits:
\alg{
&{\tilde U}_A:=\sum_{k,l=0,1}\proj{k,l}^{A'A''}\!\otimes(\sigma_x^{k}\sigma_z^{l})^A,\nn\\
&{\tilde U}_B:=\sum_{m,n=0,1}\proj{m,n}^{B'B''}\!\otimes(\sigma_x^{m}\sigma_z^{n})^B,\nn\\
&{\tilde V}_E:=\sum_{r,s=0,1}\proj{r,s}^{E'E''}\otimes(\sigma_x^{r}\sigma_z^{s}\otimes I)^{E_AE_B}.\nn
}
Then Alice and Bob discard systems $A'A''$ and $B'B''$, and Eve performs the basis transformation $\ket{\Phi_{pq}}\rightarrow|p,q\rangle$.
\end{itemize}
Due to (\ref{eq:shigechii}), we have
\alg{
&\langle k,l,m,n|^{A'A''B'B''}({\tilde U}_A\otimes {\tilde U}_B\otimes{\tilde V}_E)\nn\\
&\quad\quad\quad\quad|{k,l,m,n,k\oplus m, l\oplus n}\rangle^{A'A''B'B''E'E''}|{\bf\Phi}\rangle^{ABE_AE_B}\nn\\
&=|k\oplus m, l\oplus n\rangle^{E'E''}\nn\\
&\quad\otimes((\sigma_x^{k}\sigma_z^{l})^A\otimes(\sigma_x^{m}\sigma_z^{n})^B\otimes(\sigma_x^{k\oplus m}\sigma_z^{l\oplus n})^{E_A}\otimes I^{E_B})|{\bf\Phi}\rangle\nn\\
&=|k\oplus m, l\oplus n\rangle^{E'E''}\nn\\
&\quad\quad\quad\otimes\frac{1}{2}\sum_{p,q=0,1}(-1)^{np\oplus mq\oplus n(k\oplus m)}|\Phi_{pq}\rangle^{AB}|\Phi_{pq}\rangle^{E_AE_B}\label{eq:raashi}
}
for each value of $k$, $l$, $m$ and $n$. Noting that $\Psi_{\rm I\!I}^{*A'B'E'}\otimes\Psi_{\rm I\!I}^{*A''B''E''}$ is represented as
\alg{
&\Psi_{\rm I\!I}^{*A'B'E'}=\frac{1}{16}\sum_{k,l,m,n=0,1}\proj{k,l}^{A'A''}\otimes\proj{m,n}^{B'B''}\nn\\
&\quad\quad\quad\quad\quad\quad\quad\quad\otimes\proj{k\oplus m,l\oplus n}^{E'E''},\nn
}
the state after discarding of $A'A''$ and $B'B''$ by Alice and Bob is equal to the probabilistic mixture of (\ref{eq:raashi}) over $k,l,m,n\in\{0,1\}$ with respect to the uniform distribution, which is given by
\alg{
\!\Psi_{\rm I\!I\!I}'^{ABE_AE_B}:=\frac{1}{4}\sum_{p,q=0,1}|\Phi_{pq}\rangle\!\langle\Phi_{pq}|^{AB}\otimes|\Phi_{pq}\rangle\!\langle\Phi_{pq}|^{E_AE_B}.\nn
}
It is straightforward to verify that the above state is transformed to $\Psi_{\rm I\!I\!I}$ by the basis transformation $\ket{\Phi_{pq}}\rightarrow|p,q\rangle$.

Suppose Alice flips two fair coins, records the results $k$ and $l$ on her register, and performs $\sigma_x^k\sigma_z^l$ on her qubit. In the same way, Bob performs $\sigma_x^m\sigma_z^n$ on his qubit depending on the result of two coin flips $m$ and $n$, which is recorded on his register. Due to \req{irrurura}, the state obtained from $\Phi_{\rm I\!I\!I}$ by this protocol is given by
\alg{
&\frac{1}{4}\sum_{p,q=0,1}(\sigma_x^k\sigma_z^l\otimes \sigma_x^m\sigma_z^n)|\Phi_{pq}\rangle\!\langle\Phi_{pq}|(\sigma_x^k\sigma_z^l\otimes \sigma_x^m\sigma_z^n)^{\dagger AB}\nn\\
&\quad\quad\quad\quad\quad\quad\otimes\proj{p,q}^E\nn\\
&=\frac{1}{4}\sum_{p,q=0,1}(\sigma_x^p\sigma_z^q\otimes I)|\Phi_{k\oplus m,l\oplus n}\rangle\!\langle\Phi_{k\oplus m,l\oplus n}|(\sigma_x^p\sigma_z^q\otimes I)^{\dagger AB}\nn\\
&\quad\quad\quad\quad\quad\quad\otimes\proj{p,q}^E\nn
}
for each value of $k$, $l$, $m$ and $n$. After that, they send their qubits to Eve, who subsequently performs the controlled operation
\alg{
V^{ABE}:=\sum_{p,q=0,1}(\sigma_x^p\sigma_z^q)^A\otimes I^B\otimes\proj{p,q}^E\nn
}
to obtain the state
\alg{
|\Phi_{k\oplus m,l\oplus n}\rangle\!\langle\Phi_{k\oplus m,l\oplus n}|^{AB}\otimes\frac{1}{4}\sum_{p,q=0,1}\proj{p,q}^E\nn
}
for each $k$, $l$, $m$ and $n$. By Eve locally performs the basis transformation $|\Phi_{k\oplus m,l\oplus n}\rangle\rightarrow\ket{k\oplus m}\ket{l\oplus n}$, the state $\Psi_{\rm I\!I}^{*\otimes 2}$ is obtained.

\hfill

\section{Proof of (\ref{eq:phiinsep})-(\ref{eq:roki}) and Table \ref{tb:values}}\label{app:increl}

In this appendix, we prove Relations (\ref{eq:phiinsep})-(\ref{eq:roki}) and compute the values of non-Markovianity monotones of the states defined in Section \ref{sec:unitres}, which are listed in Table \ref{tb:values}. We denote by $\pi_2$ the maximally mixed state on a qubit system, i.e., $\pi_2:=(\proj{0}+\proj{1})/2$.

\subsection{$\Phi_{\rm I}$}
Since $\Phi_{\rm I}^{AB}=\proj{\Phi_2}^{AB}$, it is straightforward to verify that $\Phi_{\rm I}\in{\ca S}_{\rm all}\backslash{\ca S}_{\rm sep}$, and that any extension $\varrho^{ABQQ_1Q_2}$ of $\Phi_{\rm I}^{AB}$ is written in the form of
\alg{
\varrho^{ABQQ_1Q_2}=\proj{\Phi_2}^{AB}\otm\varsigma^{QQ_1Q_2}.\nn
}
Hence we have
\alg{
I(AQ_1:BQ_2|Q)_\varrho&=I(A:B|Q)_{\Phi_2}+I(Q_1:Q_2|Q)_\varrho\nn\\
&=2+I(Q_1:Q_2|Q)_\varrho\laeq{lisbon3}\\
&\geq2.\nn
}
Suppose $Q_1=Q_2=\emptyset$ and consider the correspondences
\alg{
&Q=E,\;\varsigma=\proj{0},\nn\\
&Q=E,\;\varsigma={\ca T}(\proj{0}),\nn\\
&Q=F,\;\varsigma=\rho,\nn\\
&Q=R,\;\varsigma=\rho',\nn
}
where $\rho$ and $\rho'$ are arbitrary states on $F$ and $R$, respectively, and $\ca T$ is an arbitrary operation on $E$. For each of these correspondences, Equality \req{lisbon3} implies $I(A:B|Q)_\varrho=2$. By taking the infimum over all $\rho$, $\rho'$ and $\ca T$, we obtain
\alg{
&I_M(\Phi_{\rm I})=I_\downarrow(\Phi_{\rm I})=I_\downarrow^*(\Phi_{\rm I})=I_{sq}(\Phi_{\rm I})=2.\nn
}
Under the correspondences
\alg{
&Q=E,Q_1=F_A,Q_2=F_B,\nn\\
&\quad\quad\quad\varrho={\ca T}(\proj{0})\otm\proj{0}^{F_A}\otm\proj{0}^{F_B}\nn\\
&Q=F,Q_1=E_A,Q_2=E_B,\nn\\
&\quad\quad\quad\varrho=\proj{0}^F\otm{\ca W}(\proj{0})^{E_AE_B}\nn,
}
where ${\ca W}$ is a linear isometry from $E$ to $E_AE_B$ defined by ${\ca W}(\proj{0})=\proj{0}^{E_A}\otm\proj{0}^{E_B}$, Equality \req{lisbon3} implies $I(AQ_1:BQ_2|Q)_\varrho=2$. Thus we obtain
\alg{
J_\downarrow(\Phi_{\rm I})=J_\downarrow^*(\Phi_{\rm I})=2.\nn
}

\subsection{$\Phi_{\rm I\!I}$}

Observe that for any ${\ca T}\in{\mathbb L}_E$ we have
\alg{
{\ca T}(\Phi_{\rm I\!I})=\frac{1}{2}\left(|\Phi_{00}\rangle\!\langle\Phi_{00}|^{AB}\otimes\tau_{0}^E+|\Phi_{01}\rangle\!\langle\Phi_{01}|^{AB}\otimes\tau_{1}^E\right),\nn
}
where $\tau_0:={\ca T}(\proj{0})$ and $\tau_1:={\ca T}(\proj{1})$. Denoting $(\tau_0+\tau_1)/2$ by ${\bar\tau}$, this leads to
\alg{
{\ca T}(\Phi_{\rm I\!I})^{AE}=\pi_2^A\otimes{\bar\tau}^E,\;{\ca T}(\Phi_{\rm I\!I})^{BE}=\pi_2^B\otimes{\bar\tau}^E.\nn
}
Hence we have 
\alg{
I_M({\ca T}(\Phi_{\rm I\!I}))
&=S(AE)_{{\ca T}(\Phi_{\rm I\!I})}+S(BE)_{{\ca T}(\Phi_{\rm I\!I})}\nn\\
&\quad\quad-S(E)_{{\ca T}(\Phi_{\rm I\!I})}-S(ABE)_{{\ca T}(\Phi_{\rm I\!I})}\nn\\
&=2+2S(\bar\tau)-S(\bar\tau)-1-\frac{S(\tau_0)+S(\tau_1)}{2}\nn\\
&=1+S(\bar\tau)-\frac{S(\tau_0)+S(\tau_1)}{2}\geq1,\nn
}
where the last line follows from the concavity of the von Neumann entropy. The equality holds e.g. when $\tau_0=\tau_1=\tau_2$. Hence we have
\alg{
I_\downarrow(\Phi_{\rm I\!I})=\inf_{\ca T}I_M({\ca T}(\Phi_{\rm I\!I}))=1,\nn
}
which implies $\Phi_{\rm I\!I}\notin{\ca S}_I$ due to (\ref{eq:idasmi}). Observe that a purification of $\Phi_{\rm I\!I}$ is given by
\alg{
|\phi_{\rm I\!I}\rangle^{ABEF}=\frac{1}{\sqrt{2}}\left(|\Phi_{00}\rangle^{AB}\ket{00}^{EF}+|\Phi_{01}\rangle^{AB}\ket{11}^{EF}\right),\nn
} 
and that it satisfies ${\ca U}_{\rm SWAP}^{EF}(\phi_{\rm I\!I})=\phi_{\rm I\!I}$. Hence, due to \rPrp{pfeqmonot}, we obtain $I_\downarrow^*(\Phi_{\rm I\!I})=1$ and $\Phi_{\rm I\!I}\notin{\ca S}_I^*$. It is straightforward to verify that
\alg{
\Phi_{\rm I\!I}^{AB}=\frac{1}{2}\left(\proj{00}+\proj{11}\right)\nn
}
is separable, leading to $\Phi_{\rm I\!I}\in{\ca S}_{\rm sep}$ and $I_{sq}(\Phi_{\rm I\!I})=0$. Hence we obtain $\Phi_{\rm I\!I}\in{\ca S}_{\rm sep}\backslash{\ca S}_I\cup{\ca S}_I^*$.

Let $\ca T$ be an arbitrary operation on $E$, $\ca W$ be an arbitrary linear isometry from $F$ to $F_AF_B$, and define
\alg{
&\rho_1^{ABFE}:={\ca T}(|\phi_{\rm I\!I}\rangle\!\langle\phi_{\rm I\!I}|^{ABEF}),\nn\\
&\rho_2^{ABF_AF_BE}:=({\ca T}\otimes{\ca U})(|\phi_{\rm I\!I}\rangle\!\langle\phi_{\rm I\!I}|^{ABEF}).\nn
}
For the above states, we have
\alg{
&I(AF_A:BF_B|E)_{\rho_2}\nn\\
=&I(F_A:BF_B|E)_{\rho_2}+I(A:BF_B|F_AE)_{\rho_2}\nn\\
=&I(F_A:BF_B|E)_{\rho_2}+I(A:BF_AF_B|E)_{\rho_2}\nn\\
&\quad-I(A:F_A|E)_{\rho_2}\nn\\
\geq&I(A:BF_AF_B|E)_{\rho_2}\nn\\
&\quad+I(F_A:BF_B|E)_{\rho_2}-I(F_A:AF_B|E)_{\rho_2},\label{eq:kahoko}
}
where the first two equalities follows from the chain rule of the conditional quantum mutual information and the inequality from the data processing inequality (under discarding of system $F_B$). Noting that the states $\phi_{\rm I\!I}$, $\rho_1$ and $\rho_2$ are invariant under exchanging systems $A$ and $B$, we have
\alg{
I(F_A:BF_B|E)_{\rho_2}-I(F_A:AF_B|E)_{\rho_2}=0.\nn
}
In addition, the invariance of CQMI under local isometry operations yields
\alg{
I(A:BF_AF_B|E)_{\rho_2}=I(A:BF|E)_{\rho_1}.\nn
}
Substituting the above two equalities into (\ref{eq:kahoko}) leads to
\alg{
I(AF_A:BF_B|E)_{\rho_2}\geq I(A:BF|E)_{\rho_1}.\label{eq:lion}
}
Let $\ca D$ be the dephasing operation on $F$ with respect to the computational basis, and define
\alg{
{\tilde\rho}_1^{ABFE}:={\ca D}(\rho_1^{ABFE}).\nn
}
We have
\alg{
&I(A:BF|E)_{\rho_1}\nn\\
\geq&I(A:BF|E)_{{\tilde\rho}_1}\nn\\
=&I(A:F|E)_{{\tilde\rho}_1}+I(A:B|EF)_{{\tilde\rho}_1}\nn\\
\geq&I(A:B|EF)_{{\tilde\rho}_1}\nn\\
=&I(A:BE|F)_{{\tilde\rho}_1}-I(A:E|F)_{{\tilde\rho}_1}\nn\\
=&I(A:BE|F)_{{\tilde\rho}_1}\nn\\
\geq&I(A:B|F)_{{\tilde\rho}_1}\nn\\
=&2,\label{eq:kaeru}
}
where the second and the seventh lines follow from the data processing inequality; the third and the fifth lines from the chain rule; and the sixth and the last lines from the fact that we have
\alg{
{\tilde\rho}_1^{AEF}&=\pi_2^A\otimes\frac{1}{2}(\tau_0^{E}\otimes\proj{0}^F+\tau_1^{E}\otimes\proj{1}^F),\nn\\
{\tilde\rho}_1^{ABF}&=\frac{1}{2}(\proj{\Phi_{00}}^{AB}\otimes\proj{0}^F\nn\\
&\quad\quad\quad\quad+\proj{\Phi_{01}}^{AB}\otimes\proj{1}^F).\nn
}
Combining (\ref{eq:lion}) and (\ref{eq:kaeru}), we obtain
\alg{
I(AF_A:BF_B|E)_{\rho_2}\geq2,\nn
}
which implies $J_\downarrow(\rho)\geq2$ by taking the infimum over all $\ca T$ and $\ca W$. It is straightforward to verify that the equality holds when $\ca T$ is the dephasing operation on $E$ with respect to the computational basis and $\ca U$ is a linear isometry such that $\ket{0}\rightarrow\ket{0}\ket{0}$ and $\ket{1}\rightarrow\ket{1}\ket{0}$. Applying \rPrp{pfeqmonot} also yields $J_\downarrow^*(\rho)=2$.

\subsection{$\Phi_{\rm I\!I\!I}$}

First, from $I(A:B|E)_{\Phi_{\rm I\!I\!I}}=2$, we have $\Phi_{\rm I\!I\!I}\notin{\ca S}_{\rm Markov}$. To prove $\Phi_{\rm I\!I\!I}\in{\ca S}_J$, consider purifications of $\Phi_{\rm I\!I\!I}$ defined by
\alg{
&\ket{\phi_{\rm I\!I\!I}}^{ABEF}:=\:\frac{1}{2}\sum_{p,q=0,1}\ket{\Phi_{pq}}^{AB}\ket{p,q}^E\ket{p,q}^{F},\nn\\
&\ket{\phi_{\rm I\!I\!I}'}^{ABEF_AF_B}:=\:\frac{1}{2}\sum_{p,q=0,1}\ket{\Phi_{pq}}^{AB}\ket{\Phi_{pq}}^{F_AF_B}\ket{p,q}^E.\nn
}
Define an orthonormal basis $\{\ket{e_{r,s}}\}_{r,s=0,1}$ by 
\alg{
\ket{e_{r,s}}=\frac{1}{2}\sum_{p,q=0,1}(-1)^{rp\opl sq}\ket{p,q}.\nn
}
Applying \req{furu} and \req{furuu}, we have
\alg{
\bra{e_{r,s}}^E\ket{\phi_{\rm I\!I\!I}'}^{ABEF}&=\:\frac{1}{4}\sum_{p,q=0,1}(-1)^{rp\opl sq}\ket{\Phi_{pq}}^{AB}\ket{\Phi_{pq}}^{F_AF_B}\nn\\
&=\:\frac{1}{2}\ket{\Phi_{rs}}^{AF_A}\ket{\Phi_{rs}}^{BF_B}.\nn
}
Therefore, with $\ca T$ being an operation on $E$ defined by
\alg{
{\ca T}(\cdot)=\sum_{r,s=0,1}\proj{e_r,e_s}(\cdot)\proj{e_r,e_s},\nn
}
we have
\alg{
{\ca T}(\proj{\phi_{\rm I\!I\!I}'})&=\:\frac{1}{4}\sum_{r,s=0,1}\proj{\Phi_{rs}}^{AF_A}\otm\proj{\Phi_{rs}}^{BF_B}\nn\\
&\quad\otm\proj{e_r,e_s}^E\;\in{\ca S}_{\rm Markov}^{AF_A:BF_B|E}.\nn
}
Hence we have $\Phi_{\rm I\!I\!I}\in{\ca S}_J$, which yields $I_\downarrow(\Phi_{\rm I\!I\!I})=J_\downarrow(\Phi_{\rm I\!I\!I})=0$ due to \req{idasmiiii} and \req{hoshino1}. Observe that we have ${\ca U}_{\rm SWAP}^{EF}(\phi_{\rm I\!I\!I})=\phi_{\rm I\!I\!I}$. Therefore, due to \rPrp{pfeqmonot}, we obtain $\Phi_{\rm I\!I\!I}\in{\ca S}_J^*$ and $I_\downarrow^*(\Phi_{\rm I\!I\!I})=J_\downarrow^*(\Phi_{\rm I\!I\!I})=0$.

\subsection{$\Psi_{\rm I}$, $\Psi_{\rm I}^*$}

Consider ``purifications'' of $\Psi_{\rm I}$ defined by
\alg{
&\ket{\psi_{\rm I}}^{ABEF}:=\ket{\Psi_{\rm I}}^{ABE}\ket{0}^{F},\nn\\
&\ket{\psi_{\rm I}'}^{ABEF_AF_B}:=\ket{\Psi_{\rm I}}^{ABE}\ket{0}^{F_A}\ket{0}^{F_B}.\nn
}
Let $\ca D$ be the dephasing operation on $E$ with respect to the basis $\{\ket{0},\ket{1}\}$. Then we have
\alg{
{\ca D}(\psi_{\rm I}')&=\frac{1}{2}\left(\proj{000}+\proj{111}\right)^{ABE}\nn\\
&\quad\otm\proj{0}^{F_A}\otm\proj{0}^{F_B}\in{\ca S}_{\rm Markov}^{AF_A:BF_B|E},\nn
}
which implies $\Psi_{\rm I}^{ABE}\in{\ca S}_J$. Due to \req{idasmiiii} and \req{hoshino1}, it follows that $I_\downarrow(\Psi_{\rm I})=J_\downarrow(\Psi_{\rm I})=0$. Since $\Psi_{\rm I}^{ABE}$ is a pure state, any extension $\varrho_{\rm I}$ of $\Psi_{\rm I}$ is in the form of $\varrho_{\rm I}^{ABEF}=\Psi_{\rm I}^{ABE}\otimes\varrho_0^{F}$. Thus 
\alg{
I(A:B|F)_{\varrho_{\rm I}}=I(A:B)_{\Psi_{\rm I}}=1,\nn
}
which implies $I_\downarrow^*(\Psi_{\rm I})=1$ by taking the infimum over all $\varrho_0$. This implies $\Psi_{\rm I}\notin{\ca S}_I^*$. For any linear isometry $\ca W$ from $E$ to $E_AE_B$, we have 
\alg{
I(AE_A:BE_B|F)_{{\ca W}(\varrho_{\rm I})}&=I(AE_A:BE_B)_{{\ca W}(\Psi_{\rm I})}\nn\\
&=2S(AE_A)_{{\ca W}(\Psi_{\rm I})}.\laeq{naikougata}
}
Noting that
\alg{
&{\rm Tr}_{BE_B}[{\ca W}(\Psi_{\rm I})]\nn\\
&\quad=\frac{1}{2}\left(\proj{0}^A\otm{\ca W}(\proj{0})^{E_A}+\proj{1}^A\otm{\ca W}(\proj{1})^{E_A}\right),\nn
}
we have
\alg{
S(AE_A)_{{\ca W}(\Psi_{\rm I})}\geq1\nn
}
with equality if $\ca W$ is a linear isometry such that $\ket{0}\rightarrow\ket{0}\ket{0}$ and $\ket{1}\rightarrow\ket{1}\ket{0}$. Therefore, by taking the infimum over all linear isometries $\ca W$ and all extensions $\varrho_{\rm I}$, we obtain from \req{naikougata} that $J_\downarrow^*(\Psi_{\rm I})=2$. 
Observe that ${\ca U}_{\rm SWAP}^{EF}(\psi_{\rm I})$ is a purification of $\Psi_{\rm I}^*$. Therefore, due to \rPrp{pfeqmonot}, we obtain $I_\downarrow^*(\Psi_{\rm I})=J_\downarrow^*(\Psi_{\rm I})=0$, $I_\downarrow(\Psi_{\rm I})=1$, $J_\downarrow(\Psi_{\rm I})=2$ and $\Psi_{\rm I}^*\in{\ca S}_J^*\backslash{\ca S}_I$.

\subsection{$\Psi_{\rm I\!I}$, $\Psi_{\rm I\!I}^*$}

First, from $I(A:B|E)_{\Psi_{\rm I\!I}}=1$, we have $\Psi_{\rm I\!I}\notin{\ca S}_{\rm Markov}$. To prove $\Psi_{\rm I\!I}\in{\tilde{\ca S}_I}$, consider purifications of $\Psi_{\rm I\!I}$ defined by
\alg{
&\ket{\psi_{\rm I\!I}}^{ABEF}:=\:\frac{1}{2}\sum_{p,q=0,1}\ket{\Phi_{pq}}^{AB}\ket{p,q}^E\ket{p}^{F},\nn\\
&\ket{\psi_{\rm I\!I}'}^{ABEF_AF_B}:=\:\frac{1}{2}\sum_{p,q=0,1}\ket{\Phi_{pq}}^{AB}\ket{p,q}^E\ket{p}^{F_A}\ket{0}^{F_B}.\nn
}
By tracing out $E$, we have
\alg{
&\psi_{\rm I\!I}'^{ABF_AF_B}:=\:\frac{1}{4}\sum_{p,q=0,1}\proj{\Phi_{pq}}^{AB}\otm\proj{p}^{F_A}\otm\proj{0}^{F_B}\nn\\
&=U_{\rm CNOT}^{F_AA}\!\left(\pi_2^{F_A}\!\otm\frac{1}{2}\!\sum_{q=0,1}\!\proj{\Phi_{0q}}^{AB}\!\otm\proj{0}^{F_B}\right)\!U_{\rm CNOT}^{\dagger F_AA}.\nn
}
Since
\alg{
\frac{1}{2}\sum_{q=0,1}\proj{\Phi_{0q}}^{AB}=\frac{1}{2}(\proj{00}+\proj{11})^{AB}\nn
}
is a separable state between $A$ and $B$, the state $\psi_{\rm I\!I}'^{ABF_AF_B}$ is also a separable state between $AF_A$ and $BF_B$. Hence we have $\Psi_{\rm I\!I}\in{\ca S}_J$. Due to \req{idasmiiii} and \req{hoshino1}, it follows that $I_\downarrow(\Psi_{\rm I})=J_\downarrow(\Psi_{\rm I})=0$. By linear isometry ${\ca W}:\ket{p,q}^E\rightarrow\ket{\Phi_{pq}}^{E_AE_B}$, the state $\ket{\psi_{\rm I\!I}}$ is transformed to
\alg{
\ket{{\tilde\psi}_{\rm I\!I}}^{ABE_AE_BF}:=\:\frac{1}{2}\sum_{p,q=0,1}\ket{\Phi_{pq}}^{AB}\ket{\Phi_{pq}}^{E_AE_B}\ket{p}^{F}.\nn
}
Define an orthonormal basis $\{\ket{e_{r}}\}_{r=0,1}$ by 
\alg{
\ket{e_{r}}=\frac{1}{\sqrt{2}}\sum_{p=0,1}(-1)^{rp}\ket{p}.\nn
}
Applying \req{furu} and \req{furuu}, we have
\alg{
\bra{e_{r}}^F\ket{{\tilde\psi}_{\rm I\!I}}^{ABE_AE_BF}\!&=\frac{1}{2\sqrt{2}}\sum_{p,q=0,1}\!(-1)^{rp}\ket{\Phi_{pq}}^{AB}\ket{\Phi_{pq}}^{E_AE_B}\nn\\
&=\:\frac{1}{\sqrt{2}}\ket{\Phi_{r0}}^{AE_A}\ket{\Phi_{r0}}^{BE_B}.\nn
}
Therefore, we have
\alg{
{\tilde\psi}_{\rm I\!I}^{ABE_AE_B}&=\sum_{r=0,1}\bra{e_{r}}^F\proj{\psi_{\rm I\!I}}^{ABE_AE_BF}\ket{e_{r}}^F\nn\\
&=\:\frac{1}{2}\sum_{r=0,1}\proj{\Phi_{r0}}^{AE_A}\otm\proj{\Phi_{r0}}^{BE_B}\nn\\
&\;\in{\ca S}_{\rm sep}^{AE_A:BE_B},\nn
}
which yields $\Psi_{\rm I\!I}\in{\ca S}_J^*$. Hence, from \req{idasmiiii} and \req{hoshino2}, we have $I_\downarrow^*(\Psi_{\rm I})=J_\downarrow^*(\Psi_{\rm I})=0$. Observe that we have ${\ca U}_{\rm SWAP}^{EF}(\psi_{\rm I\!I})$ is a purification of $\Psi_{\rm I\!I}^*$. Therefore, due to \rPrp{pfeqmonot}, we also have $\Psi_{\rm I\!I}^*\in{\ca S}_J\cap{\ca S}_J^*$ and $I_\downarrow(\Psi_{\rm I})=J_\downarrow(\Psi_{\rm I})=I_\downarrow^*(\Psi_{\rm I})=J_\downarrow^*(\Psi_{\rm I})=0$. \QED

\end{document}